\documentstyle[amsfonts,amssymb,amsmath,graphicx]{article}   

\newcommand{\dsp}{\displaystyle}

\newcommand{\qed}{\hfill $\square$}

\newtheorem{theorem}{Theorem}[section]
\newtheorem{lemma}[theorem]{Lemma}
\newtheorem{definition}[theorem]{Definition}
\newtheorem{corollary}[theorem]{Corollary}
\newtheorem{remark}[theorem]{Remark}
\newtheorem{proposition}[theorem]{Proposition}

\title{\bf Huygens' principle  
for the generalized Dirac operator in  curved spacetime 
}

\author{{\bf Karen Yagdjian}
 }

\begin{document}

\date{}
\maketitle
\thispagestyle{empty}
\vspace{-0.3cm}

\begin{center}
{\it School of Mathematical and Statistical Sciences,
University of Texas RGV,\\
1201 W.~University Drive,  
Edinburg, TX 78539,
USA \\
E-mail: karen.yagdjian@utrgv.edu}
\end{center}
\medskip

\addtocounter{section}{-1}
\renewcommand{\theequation}{\thesection.\arabic{equation}}
\setcounter{equation}{0}
\pagenumbering{arabic}
\setcounter{page}{1}
\thispagestyle{empty}

\hspace{2cm}\begin{abstract}
\begin{small}
In this article we give  sufficient conditions for the  generalized Dirac operator to obey the incomplete Huygens  principle,   as well as  necessary and sufficient  conditions to obey the Huygens  principle by the  Dirac operator in the curved spacetime of the Friedmann-Lema\^itre-Robertson-Walker models of cosmology. 
 
\medskip

\end{small}
\end{abstract}

\setcounter{equation}{0}
\renewcommand{\theequation}{\thesection.\arabic{equation}}

\section{Introduction}

An interesting question in the physics of fundamental particles (fields)   is the validity of the Huygens  principle. 
In this article we find sufficient conditions for the incomplete  Huygens  principle for   the generalized Dirac operator  and necessary conditions  for the  Dirac operator in the curved spacetime of the Friedmann-Lema\^itre-Robertson-Walker (FLRW) models of cosmology. 
 We use the definition of the Huygens  principle due to Hadamard \cite{Wunsch} as the absence of tails.  Thus,  
the field equations  
satisfy the Huygens  principle if and only if the solution has no tail, that is, solution depends
on the source distributions on the past null cone of the field only and not
on the sources inside the cone. 
\medskip

Fields of nonzero  spin  in a curved space have been studied   from the mathematical and the physical (classical and quantum) point of view (see, e.g., \cite{Fock,Parker,Schrodinger} and references therin). 
It is known that, given  Dirac equation in a curved  four-dimensional spacetime $\left(M, g_{a b}\right)$, the Huygens  principle is generally violated by its solutions, due to the  mass term in the  equation  and the curvature of spacetime \cite{Faraoni}.  The presence or absence of tails for   waves   has been established   for  some  spacetime metrics $g_{a b}$, including constant  curvature metrics \cite{Wunsch}. In fact, the study of  the Huygens  principle has important applications to quantum field theory and cosmology,  especially in inflationary theories of the early universe. 
The Huygens  principle is a   fundamental feature  of the physics and the mathematics
of  propagation of waves.
The fact that  
the support of the commutator or the anticommutator-distribution, respectively,
lies on the null-cone if and only if the Huygens  principle holds for the corresponding
 equation \cite{Lichnerowicz} shows significance of the Huygens  principle    
for quantum field theory.  
The Huygens  principle is also studied for  the gravitational waves in the  curved background (see, e.g.,   \cite{Kulczycki}). 
In the analysis of partial differential equations the   
Huygens  principle plays important role in the estimates of the Bahouri-G\'erard concentration 
compactness method and Strichartz estimates (see, e.g.,   \cite{Krieger,Schlag}) and in study of blowup of solutions of nonlinear hyperbolic equations (see, e.g.,  \cite{Alinhac}).  
\medskip

The spin-$\frac{1}{2}$ particle  field in the Minkowski spacetime (see, e.g., \cite[(20) Ch.1, Sec.6]{B-Sh})
\begin{eqnarray}
\label{0.1}
&  &  \left(i {\gamma }^0  \partial_0+i  \gamma ^k \partial_k    -m{\mathbb I}_4\right)\Psi (x,t)= 0, \quad x \in {\mathbb R}^3  , \,\, t \in {\mathbb R}_+\,, \quad m \in {\mathbb C}\,,
\end{eqnarray}
satisfies the  Huygens  principle if and only if $m=0$. Here and henceforth, Einstein summation convention over repeated indexes is employed and also the speed of light $c$ and  Planck's constant $h$   are set equal to  unity. Even if we relax the  Huygens  principle  
by considering  the Cauchy problem for (\ref{0.1}) with the initial condition
\begin{eqnarray*}
&  &
\Psi (x,0)= (\Phi_0(x ),\Phi_1(x ),\Phi_2(x ),\Phi_3(x ))^T   , \quad x \in {\mathbb R}^3  \,,   
\end{eqnarray*}
that has only one non-vanishing component, say $\Phi_1(x )=\Phi_2(x )=\Phi_3(x )=0 $,  the solution of the equation with $m \not=0 $  depends
on the values of the function $\Phi_0(x )$  inside the   past null cone of the field. 
\medskip

The metric tensor in the spatially flat FLRW spacetime in Cartesian coordinates is 
\[
(g_{\mu \nu })=    \left (
   \begin{array}{ccccc}
 1& 0& 0   & 0 \\
   0& -a^2(t) &  0 & 0 \\ 
 0 & 0 &  -a^2(t)   & 0 \\
 0& 0& 0   &  -a^2(t) \\
   \end{array} \right),\quad \mu ,\nu =0,1,2,3.  
\] 
We will focus on the de~Sitter space with the scale factor   $a(t)=e^{Ht}  $ (see, e.g., \cite{Moller}) that is modeling the expanding or contracting universe if $H>0$ or $H<0$, respectively. 
The Dirac equation in the de~Sitter space is (see, e.g., \cite{Barut-D})
\begin{equation}
\label{DE}
  \dsp 
\left(  i {\gamma }^0    \partial_0   +i e^{-Ht}{\gamma }^1  \partial_1+i  e^{-Ht}{\gamma }^2 \partial_2+i e^{-Ht}{\gamma }^3   \partial_3 +i \frac{3}{2}    H {\gamma }^0     -m{\mathbb I}_4 \right)\Psi=F \,,
\end{equation}
where $F$ is a source term, while   
the contravariant gamma matrices are   (see,  e.g., \cite[p. 61]{B-Sh})
\begin{eqnarray*}
&  &
 \gamma ^0= \left (
   \begin{array}{ccccc}
   {\mathbb I}_2& {\mathbb O}_2   \\
   {\mathbb O}_2& -{\mathbb I}_2   \\ 
   \end{array}
   \right),\quad 
\gamma ^k= \left (
   \begin{array}{ccccc}
  {\mathbb O}_2& \sigma ^k   \\
  -\sigma ^k &  {\mathbb O}_2  \\  
   \end{array}
   \right),\quad k=1,2,3\,.
 \end{eqnarray*}
Here $\sigma ^k $ are Pauli matrices 
\begin{eqnarray*}
&  &
\sigma ^1= \left (
   \begin{array}{ccccc}
  0& 1   \\
  1& 0  \\  
   \end{array}
   \right), \quad
\sigma ^2= \left (
   \begin{array}{ccccc}
  0& -i   \\
  i& 0  \\  
   \end{array}
   \right),\quad
\sigma ^3= \left (
   \begin{array}{ccccc}
  1& 0   \\
  0&-1 \\  
   \end{array}
   \right)\,,
\end{eqnarray*}
and  ${\mathbb I}_n $, ${\mathbb O}_n $ denote the $n\times n$ identity and zero matrices, respectively. We note here that for $H = 0$ and $F=0 $  the  equation (\ref{DE}) coincides with (\ref{0.1}).

The construction
of a quantum field theory in curved spacetimes
and the definition of a quantum vacuum demand  a detailed
investigation of the solutions of relativistic  equations
in curved backgrounds. (See, e.g., \cite{Birrell}.)  
  In \cite{ArX2020} author presents the fundamental solutions and the solutions to the Cauchy problem via classical formulas for the wave equation in the Minkowski spacetime and the certain integral transform involving the Gauss's hypergeometric function in the kernel.  One can regard that integral transform as an analytical mechanism that from the massless field in the Minkowski spacetime generates massive particle in the curved spacetime. 
\medskip

In the terms of the fundamental solutions the Huygens 
principle sustains that a delta-like impulse of field travels on a sharp front propagating along the light cone.
  
\medskip

 Even though nowadays, numerical solutions of   differential equations are 
available, in some cases    a deep understanding of 
 properties of the solutions  is  possible  only 
by the examination of the  explicit formulas when they are known. This is the case with the Huygens  principle. Some known results on the Huygens  principle
for the Dirac equation can be found in \cite{Chalub,Deser-Nepomechie,Faraoni,Gunter,McLenaghan-Sasse,Pascazio,Sonego-Faraoni,Wunsch}. 
\medskip

 Recall  that   a retarded fundamental solution (a retarded inverse) for the Dirac operator (\ref{DE})  is a matrix operator  $ 
{\mathcal E}^{ret}={\mathcal E}^{ret} \left(x, t ; x_{0}, t_{0};m\right)
 $    that solves the equation
\begin{eqnarray}
\label{FSE} 
\left(  i {\gamma }^0    \partial_0   +i e^{-Ht}{\gamma }^\ell  \partial_\ell   +i     \frac{3}{2}   H  {\gamma }^0 -m{\mathbb I}_4 \right){\mathcal E}  \left(x, t ; x_{0}, t_{0};m\right)
& = &\delta\left(x-x_{0}, t-t_{0}\right){\mathbb I}_4, \nonumber \\
&  &
 (x,t,x_0,t_0 ) \in {\mathbb R}^8,  
\end{eqnarray}
and with the support in the {\it causal  future}    $D_+(x_0, t_0)$  of the point $(x_0,t_0)  \in {\mathbb R}^4$. The
advanced fundamental solution (propagator) $ 
{\mathcal E}^{adv}={\mathcal E}^{adv}  (x, t ; x_{0}, t_{0};$ $  m )
 $ solves the equation (\ref{FSE}) and has a  support in the {\it causal  past}  $D_-(x_0, t_0)$.  The forward and backward light cones are defined as the boundaries of  $D_+(x_0, t_0)$ and $D_-(x_0, t_0)$, respectfully, where
\[ 
D_{\pm}\left(x_{0}, t_{0}\right) :=\left\{(x, t) \in {\mathbb R}^{3+1} ;
\left|x-x_{0}\right| \leq \pm\left(\phi (t) -\phi (t_{0})  \right)\right\}\,,
\]
and  $\phi (t):= (1-e^{-Ht} )/H$ is a distance function.  
In fact, any intersection of $D_-(x_0, t_0)$ with the hyperplane $t = const < t_0$ determines the
so-called {\it dependence domain} for the point $ (x_0, t_0)$, while the intersection of $D_+(x_0, t_0)$
with the hyperplane $t = const > t_0$ is the so-called {\it domain of influence} of the point
$ (x_0, t_0)$. 
The Dirac equation (\ref{DE})  is non-invariant with respect to time inversion and its solutions have different properties in different direction of time.
\medskip

Let ${\mathcal A}(x,\partial_x) $ be a   differential operator ${\mathcal A}(x,\partial_x)=\sum_{|\alpha| \leq p} a_\alpha (x)D_x^\alpha $ and the coefficients $ a_\alpha (x)$ 
are $C^\infty$-functions in  ${\mathbb R}^3 $, that is $a_\alpha \in C^\infty ({\mathbb R}^3) $. We consider defined in \cite{ArX2020} the generalized Dirac equation 
\begin{eqnarray}
\label{EqAA}
\left( i {\gamma }^0  \partial_0+i   e^{-Ht} \gamma ^k A_k(x, \partial_x)  +i   \frac{3}{2}  H    \gamma ^0   -m{\mathbb I}_4 \right)\Psi  =0\,,
\end{eqnarray} 
where
${\mathcal A} (x, \partial_x)$, $A_k(x, \partial_x) $, $k=1,2,3$, are the scalar operators with  the property
\begin{eqnarray}
\label{AA}
&  &
\gamma ^k  A_k(x, \partial_x)  \gamma ^j A_j(x, \partial_x)= -   {\mathcal A} (x, \partial_x){\mathbb I}_4\,.
\end{eqnarray}
For $A_k(x, \partial_x)=\partial_{x_k}$, $k=1,2,3$, and  ${\mathcal A} (x, \partial_x) = \Delta $ the   equation (\ref{EqAA})  is the Dirac equation  (\ref{DE}) without source term. Several examples of the generalized Dirac equation, including the equation for the  motion of the charged spin-$\frac{1}{2}$ particle in a 
constant homogeneous magnetic field,   one can find in \cite{ArX2020}. Denote ${\mathcal E}^{we,\mathcal A}(x,t,D_x)$ the solution operator (fundamental solution)   of the   problem
\begin{eqnarray*}
\begin{cases} 
 v_{tt}-   {\mathcal A}(x,\partial_x)  v =0, \quad x \in {\mathbb R}^3  \,,\quad t \in [0,\infty)\,, \cr
 v(x,0)= \varphi (x), \quad v_t(x,0)=0\,,\quad x \in {\mathbb R}^3 \,,  
\end{cases} 
\end{eqnarray*}
while  by $ {\mathcal E}^{we,\mathcal A}(x,t,y)$ we denote the Schwartz kernel of  ${\mathcal E}^{we,\mathcal A}(x,t,D_x)$.

The so-called incomplete   Huygens  principle for the Klein-Gordon equation in the de~Sitter spacetime was introduced in \cite{JMP2013}. Here we define corresponding principle for the   Dirac equation in the de~Sitter spacetime.
\begin{definition}
We say that the equation (\ref{EqAA}) obeys the  incomplete   Huygens  principle with respect to the first 2-spinor  initial data $ 
\Phi _0, \Phi _1 $   if the solution $\Psi =(\Psi_0,\Psi _1,\Psi _2,\Psi _3)^t  $ with the second 2-spinor data  $ 
\Phi _2=\Phi _3=0 $  vanishes at all points which cannot be
reached from the support of initial data by a null geodesic.

We say that the equation (\ref{EqAA}) obeys the  incomplete   Huygens  principle with respect to the second 2-spinor  initial data $ 
\Phi _2, \Phi _3 $   if the solution $\Psi =(\Psi_0,\Psi _1,\Psi _2,\Psi _3)^t  $ with the first  2-spinor data  
$ \Phi _0=\Phi _1=0 $  vanishes at all points which cannot be
reached from the support of initial data by a null geodesic.
\end{definition}

It is evident that if the equation obeys the  Huygens  principle, then it obeys the   incomplete Huygens  principle.  The main result of this paper is the following theorem. 
\begin{theorem}
\label{T0.3}
Consider the generalized Dirac equation (\ref{EqAA}).\\
\noindent
\mbox{\rm (i)} If  $m=0 $ and the fundamental solution ${\mathcal E}^{we,\mathcal A}(x,t,D_x)$  with the differential operator ${\mathcal A}(x,\partial_x) $   obeys the Huygens  principle, then  the generalized Dirac equation (\ref{EqAA})  obeys it as well. 
\medskip

\noindent
\mbox{\rm (ii)} If  $m=iH $ and the fundamental solution ${\mathcal E}^{we,\mathcal A}(x,t,D_x)$  with the differential operator ${\mathcal A}(x,\partial_x) $   obeys the Huygens  principle, then  the generalized Dirac equation (\ref{EqAA})   
obeys the incomplete Huygens  principle with respect to the first 2-spinor  initial data $ 
\Phi _0, \Phi _1 $. 
\medskip

\noindent
\mbox{\rm (iii)} If  $m=-iH $ and the fundamental solution ${\mathcal E}^{we,\mathcal A}(x,t,D_x)$  with the differential operator ${\mathcal A}(x,\partial_x) $   obeys the Huygens  principle, then  the generalized Dirac equation (\ref{EqAA}) 
obeys the incomplete Huygens  principle with respect to the second 2-spinor  initial data $ 
\Phi _2, \Phi _3$.
\medskip

\noindent
\mbox{\rm (iv)} The solution of the Dirac equation (\ref{DE}) with the mass $m \in {\mathbb C}$   in the de~Sitter spacetime obeys the  incomplete Huygens  principle with respect to the first 2-spinor  initial data $\Phi _0, \Phi _1 $ if and only if $m=0$ or $m=iH$. 
\medskip

\noindent
\mbox{\rm (v)} The solution of the Dirac equation (\ref{DE}) with the mass $m \in {\mathbb C}$   in the de~Sitter spacetime obeys the  incomplete Huygens  principle with respect to the second 2-spinor  initial data $ 
\Phi _2, \Phi _3$ if and only if $m=0$ or $m= -iH$.   
\medskip

\noindent
\mbox{\rm (vi)} The solution of the Dirac equation (\ref{DE}) with the mass $m \in {\mathbb C}$   in the de~Sitter spacetime obeys the Huygens  principle if and only if $m=0$ or $m= iH,-iH $.  
\end{theorem} 

\begin{remark}
We do not know if the generalized Dirac equation is   necessarily produced by a spacetime metric. 
\end{remark}

\begin{remark}
The statement \mbox{\rm (vi)} follows from Theorem by Wunsch~\cite{Wunsch}. In the present paper we give another proof based on the explicit formula for the solution. 
\end{remark}

As it is proved by Wunsch~\cite{Wunsch} the system for symmetric spinor fields with the spin $s=\frac{1}{2}(n+1)$ ($n=0,1,2,\ldots)$ in four dimensional spacetime satisfies the Huygens  principle if and only if the spacetime has  constant curvature. 
The de~Sitter spacetime has a constant curvature. 
\medskip

For the  particle field satisfying the Dirac equation (\ref{DE}) in the de~Sitter spacetime (in FLRW model)   according to Theorem~\ref{T0.3}, there are three values of mass with which the equation  obeys the  Huygens  principle. One of them  $m=0$, two others are $ m=iH$ and $m=-iH$.  The last two appear due to the curvature of the spacetime $R=-12H^2\not= 0$, where the Hubble constant $H$ is very small number around $2.2\times 10^{-18}s^{-1} $. The particle with $m=0$ in the standard textbooks refers as {\it neutrino} \cite{B-Sh}. It is now known that there are three discrete neutrino masses with different tiny values less than   $ 2.14\times 10^{-34}$ grams.  It is very tempting to think that these three masses $ m=0,iH,-iH$ correspond to three  neutrinos. But possible barrier can be the fact that  the  mass $m=\pm iH$ is an imaginary number. If $m$ of the Dirac equation (\ref{DE}) is responsible for the gravitational interaction  according to Newton's law,  then it will be repulsive for two particles with the imaginary mass $m=iH$ and also between two particles with the imaginary mass $m=-iH$. On the other hand, it is commonly accepted that neutrino moves with speed of light. For interaction of neutrinos and gravitational fields see \cite{Brill-Wh}. In particular, in \cite{Brill-Wh} is considered the stress-energy tensor of the neutrino field, which must be inserted in the Einstein's equations, if the reaction of the neutrino field on the gravitational field is to be described.

It is possible that the   Huygens  principle affects the ability of particle to interact with other fields.  The mass of a stationary electron  has the  value of about $9.1094\times  10^{-28} g$. For the massive particle obeying the   Huygens  principle ($h\approx 6.6261 \times  10^{-27} cm^2\cdot  g \cdot s^{-1}$, $c\approx 2.99792458 \times  10^{10} cm \cdot s^{-1}$) in the physical units 
\[
|m|=\frac{Hh}{c^2} 
\approx 4.9\times 10^{-65} g\,.
\]
The duration of time  when the factor $\exp \left(   - i \,m c^2 t/h\right)  $ of the kernel $K_1 $~(\ref{K1new}) 
of the integral representing   the solution  of the generalized Dirac equation  with the mass $m=\pm i Hh/c^2$  stays around unity   is limited by 
\[ 
T\approx \frac{h}{c^2m} \approx  10^{18}sec\,.
\]
The age of the universe is $ \approx  10^{18}$ sec. The only electrically neutral and long-lived particles in the Standard Model  of particle physics are the neutrinos \cite{Boyarsky}.  
\medskip

As experiments show that neutrinos have mass and they  in principle are a very natural Dark Matter candidate. The reasons why the known neutrinos cannot compose all of the observed Dark Matter are the smallness of their mass and the magnitude of their coupling to other particles. 
Despite their tiny masses,  neutrinos  are so numerous that their gravitational force can influence other matter in the universe. 
The  weak interactions of neutrinos make it very difficult to study their properties. 
The observed neutrino flavour oscillations clearly indicate that {\it at least two neutrinos have non-vanishing mass}  \cite{Boyarsky}. 
\medskip

Since the neutrino does not respond directly to electromagnetic  fields, if one intends to influence its
trajectory  one has to
make use of gravitational fields. In other words, one
has to consider the physics of a neutrino as a spinor in a curved
metric. 
\medskip

\medskip

From Theorem~\ref{T0.3} by letting $H \to 0$ one can learn that the spin-$\frac{1}{2}$ particle  field with mass $m \in  {\mathbb C}$ in the Minkowski spacetime obeys  the    Huygens  principle if and only if $m=0$. On the other hand,  the direct calculation  of the limit  of the kernel     $K_1(r,t;M)$ and the corresponding operator   is a challenging  exercise   although result is known  (see, e.g., \cite{ArX2020}).  
\medskip

This paper is organized as follows.  In Section~\ref{S1} we give the representation of the solution of the  generalized Dirac equation in the de~Sitter spacetime.  Then,
 in Section \ref{S2}, we prove the sufficiency part of Theorem~\ref{T0.3}. All remaining sections are devoted to the necessity part  of Theorem~\ref{T0.3}.

\section{Representation of the solution of the generalized Dirac equation in de~Sitter spacetime} 
\label{S1}

For $(x_0, t_0) \in {\mathbb R}^n\times {\mathbb R}$, $M \in {\mathbb C}$, $r = |x- x_0 | /H $, we define the function 
\begin{eqnarray}
\label{0.6} 
E(r,t;0,t_0;M)
& :=  &
4^{-\frac{M}{H}} e^{ M  (  t_0+ t)} \left(\left(e^{-H t_0}+e^{-H t}\right)^2-(H r)^2\right)^{\frac{M}{H}-\frac{1}{2}}  \nonumber \\
&  &
\times  F \left(\frac{1}{2}-\frac{M}{H},\frac{1}{2}-\frac{M}{H};1;\frac{\left(e^{-H t}- e^{-H t_0}\right)^2-(r H)^2}{\left(e^{-H t}+e^{-Ht_0}\right)^2-(r H)^2}\right)  \,,
\end{eqnarray}
where   $(x,t) \in D_+ (x_0,t_0)\cup D_- (x_0,t_0) $  and $F\big(a, b;c; \zeta \big) $ is the hypergeometric function. 
Denote  
\[
M_+=   \frac{1}{2}H  + im    ,\quad M_-=   \frac{1}{2}H  - im \,.
\]
Let $e^{H\cdot } $  be  the operator multiplication by $e^{Ht } $. 
Theorem~0.2~\cite{ArX2020}  gives the representation formula  for the solutions of the Cauchy problem. In order to formulate it we need the operator ${\cal G}(x,t,D_x;M ) $ defined by 
\[
{\cal G}(x,t,D_x;M )[f]  
=
2   \int_{ 0}^{t} db
  \int_{ 0}^{\phi (t)- \phi (b)}  E(r,t;0,b;M) \int_{{\mathbb R}^n} {\mathcal E}^w (x-y,r) f  (y,b) \,d y  \, dr  
, \quad  f \in C_0^\infty({\mathbb R}^{n+1})\,,
\]
where ${\mathcal E}^w (x,r) $ is a fundamental solution of the Cauchy problem
\begin{eqnarray*}
\begin{cases} 
 v_{tt}-  \Delta  v =0, \quad x \in {\mathbb R}^n  \,,\quad t \in {\mathbb R}\,, \cr
 v(x,0)= \varphi (x), \quad v_t(x,0)=0\,,\quad x \in {\mathbb R}^n \,,  
\end{cases}
\end{eqnarray*}
 in the Minkowski spacetime, that is, for $n=3$ 
  (see, e.g., \cite{Shatah})  
\[ 
{\mathcal E}^{w}(x, t) :=\frac{1}{4\pi } \frac{\partial}{\partial t}
\frac{1}{t} \delta(|x|-t)\,.
\]
The distribution $\delta(|x|-t)$ is defined
 by
\[
\langle \delta(|\cdot|-t), \psi(\cdot)\rangle  =\int_{|x|=t} \psi(x)\, d x \quad \mbox{\rm for } \quad \psi \in C_{0}^{\infty}\left({\mathbb R}^{3}\right).
\] 
We need also the kernel function
\begin{eqnarray}
\label{K1new}
K_1(r,t;M)
& :=  &
E(r,t;0,0;M)   
\end{eqnarray}
and the operator ${\cal K}_1(x,t,D_x;M)$, which is  defined  as follows:
\begin{equation}
\label{K1OPER}
{\cal K}_1(x,t,D_x;M) \varphi (x) = 
 2\int_{0}^{\phi (t) } 
  K_1( s,t;M)  \int_{{\mathbb R}^n} {\mathcal E}^w (x-y,s)  \varphi  (y) \,d y\, ds\,, \quad \varphi \in C_0^\infty({\mathbb R}^n).
\end{equation}

According to Theorem~0.2~\cite{ArX2020}, 
 solution to the Cauchy problem
\begin{eqnarray*}
&  &
\begin{cases} \dsp \left(i {\gamma }^0  \partial_0+i  e^{-Ht}{\gamma }^k \partial_k  + i       \frac{3}{2}  H  {\gamma }^0   -m{\mathbb I}_4\right)\Psi (x,t)=F(x,t)\,,\cr 
\Psi (x,0)= \Phi   (x ) \,,
\end{cases}
\end{eqnarray*}
$m \in {\mathbb C}$, is given by the following formula
\begin{eqnarray*} 
\Psi (x,t) 
& = &
 - e^{-Ht}\left( i\gamma ^0 \partial_0+  ie^{-Ht} \gamma ^k \partial_k- i\frac{H}{2}\gamma ^0+m{\mathbb I}_4\right) \\
&  &
\times \left [   \left (
   \begin{array}{cccc}
 {\cal G}(x,t,D_x;M_+){\mathbb I}_2& {\mathbb O}_2   \\ 
  {\mathbb O}_2 & {\cal G}(x,t,D_x;M_-){\mathbb I}_2   \\ 
   \end{array}
   \right )[e^{H\cdot }F ] \right.\\
&  &
\left. +i\gamma ^0 \left (
   \begin{array}{cccc}
 {\cal K}_1(x,t,D_x;M_+){\mathbb I}_2 & {\mathbb O}_2  \\ 
 {\mathbb O}_2&  {\cal K}_1(x,t,D_x;M_-){\mathbb I}_2   \\ 
   \end{array}
   \right )[\Phi   ]\right ]  \,.
 \end{eqnarray*}

In addition  to the functions (\ref{0.6}), (\ref{K1new})
for $M \in {\mathbb C}$  we recall one more kernel function  from \cite{Yag_Galst_CMP,MN2015}
\begin{eqnarray*}
K_0(r,t;M)
&  := &
-\left[\frac{\partial}{\partial b} E(r, t ; 0, b ; M)\right]_{b=0}\,. 
\end{eqnarray*}
Then according to \cite{MN2015} the solution operator for the Cauchy problem for the scalar {\it generalized Klein-Gordon equation}
in the de~Sitter spacetime  
\begin{eqnarray*}
&  &
\left( \partial_0^2     
-  e^{-2Ht}    {\mathcal A}(x,\partial_x)     
 -  M ^2   \right)\psi  =f,\quad \psi (x,0)= \varphi _0 (x), \quad \psi_t (x,0)= \varphi _0 (x) \,,  
 \end{eqnarray*}
where $f$ is a source term, is given as follows
\[
\psi (x,t) = {\cal G} ( x,t,D_x;M)[f]+{\cal K}_0(x,t,D_x;M)[\varphi _0 ]+ {\cal K}_1(x,t,D_x;M)[\varphi _1 ] \,.
\]

To describe the operators ${\mathcal G},  {\mathcal K}_0, {\mathcal K}_1 $ we 
  recall the results of  Theorem~1.1~\cite{MN2015}.  
For $ f \in C ({\mathbb R}^3 \times I  )$,\, $ I=[0,T]$, $0< T \leq \infty$, and \, $ \varphi_0 $,  $ \varphi_1 \in C ({\mathbb R}^3  ) $,
let  the function\,
$v_f(x,t;b) \in C_{x,t,b}^{m,2,0}({\mathbb R}^3  \times [0,(1-e^{-HT})/H]\times I)$\,
be a solution to the   problem
\begin{equation}
\label{1.22}
\begin{cases} 
 v_{tt} -   {\mathcal A}(x,\partial_x)  v  =  0 \,, \quad x \in {\mathbb R}^3  \,,\quad t \in [0,(1-e^{-HT})/H]\,,\cr
v(x,0;b)=f(x,b)\,, \quad v_t(x,0;b)= 0\,, \quad b \in I,\quad x \in {\mathbb R}^3 \,, 
\end{cases}
\end{equation}
and the function \, $  v_\varphi(x, t) \in C_{x,t}^{m,2}(\Omega \times [0,(1-e^{-HT})/H])$ \, be a   solution   of the   problem
\begin{equation}
\label{1.23}
\begin{cases} 
 v_{tt}-   {\mathcal A}(x,\partial_x)  v =0, \quad x \in {\mathbb R}^3  \,,\quad t \in [0,(1-e^{-HT})/H]\,, \cr
 v(x,0)= \varphi (x), \quad v_t(x,0)=0\,,\quad x \in {\mathbb R}^3 \,.  
\end{cases} 
\end{equation}
Then the function  $u= u(x,t)$   defined by
\begin{eqnarray*}
u(x,t)
&  =  &
2   \int_{ 0}^{t} db
  \int_{ 0}^{\phi (t)- \phi (b)}  E(r,t;0,b;M)  v_f(x,r ;b) \, dr  
+ e ^{\frac{1}{2}Ht} v_{\varphi_0}  (x, \phi (t))\\
&  &
+ \, 2\int_{ 0}^{\phi (t)}  K_0( s,t;M)v_{\varphi_0}  (x, s)   ds  \nonumber 
+\, 2\int_{0}^{\phi (t) }  v_{\varphi _1 } (x,  s)
  K_1( s,t;M)   ds
, \quad x \in {\mathbb R}^3   , \,  t \in I ,
\end{eqnarray*}
where $\phi (t):= (1-e^{-Ht} )/H$,
 solves the problem
\[
\begin{cases}
u_{tt} - e^{-2Ht}{\mathcal A}(x,\partial_x)  u - M^2 u= f, \quad  x \in {\mathbb R}^3  \,,\,\, t \in I,\cr
  u(x,0)= \varphi_0 (x)\, , \quad u_t(x,0)=\varphi_1 (x),\quad x \in {\mathbb R}^3 \,.
  \end{cases}
\] 

These representation formulas can be used for the functions in the Sobolev spaces as well. 
We stress here that the existence of the solutions in the problems (\ref{1.22}) and (\ref{1.23})  is assumed.

There exist two important examples of generalized Klein-Gordon equation. The first one has ${\mathcal A}(x,\partial_x) =\Delta  $ and it is related to the problem written in the Cartesian coordinates. (See, e.g., \cite{ArX2020}).  The second one has the equation written in the spherical coordinates $( r ,   \theta, \phi)  $. In the  FLRW  spacetime with the  line element
\begin{equation*}
ds^2=    dt^2
- e^{2Ht}\left( \frac{1}{1-Kr^2} dr^2 + r^2(d\theta ^2 + \sin^2 \theta \, d\phi ^2) \right)
\end{equation*} 
the Klein-Gordon equation is
\begin{eqnarray*}
&   &
 \partial_t ^2    \psi
+3 H\partial_ t \psi
  -  e^{-2Ht}{\mathcal A}(x,\partial_x)  \psi  + m^2 \psi =0\,,
\end{eqnarray*}
where  
\begin{eqnarray*}
{\mathcal A}(x,\partial_x) 
&  :=   &
\frac{\sqrt{1-Kr^2}}{ r^2 }\frac{\partial }{\partial r}\left(  r^2  \sqrt{1-Kr^2}\frac{\partial \psi }{\partial  r} \right)
+   \frac{1}{ r^2\sin  \theta}\frac{\partial }{\partial \theta }\left(    \sin  \theta  \frac{\partial \psi }{\partial \theta } \right)
+   \frac{1 }{ r^2\sin^2  \theta}\frac{\partial }{\partial \phi }\left( \frac{\partial \psi }{\partial \phi  } \right)
\end{eqnarray*}
is the Laplace-Beltrami operator in the  spatial variables, where $K=-1, 0$, or $+1$, for a hyperbolic, flat, or spherical spatial geometry, respectively.  For more examples see references in \cite{ArX2020}.

\medskip

Theorem~1.3~\cite{ArX2020} allows us to write solution of the  {\it generalized Dirac equation}.  More exactly, 
assume that ${\mathcal A} (x, \partial_x)$, $A_k(x, \partial_x) $, $k=1,2,3$, are the scalar operators with  the properties
(\ref{AA}). 
Then the solution to the Cauchy problem
\begin{eqnarray*}
&  &
\begin{cases} \dsp \left(i {\gamma }^0  \partial_0+i e^{-Ht} \gamma ^k A_k(x, \partial_x)  + i       \frac{3}{2}  H  {\gamma }^0   -m{\mathbb I}_4\right)\Psi (x,t)=F(x,t),\cr 
\Psi (x,0)= \Phi   (x ) 
\end{cases}
\end{eqnarray*}
is given as follows
\begin{eqnarray} 
\label{solution}
\Psi (x,t) 
& = &
- e^{-Ht}\left( i\gamma ^0 \partial_0+  ie^{-Ht} \gamma ^k A_k(x, \partial_x)- i\frac{H}{2}\gamma ^0+m{\mathbb I}_4\right) \nonumber \\
&  &
\times \left [  \left (
   \begin{array}{cccc}
 {\cal G}(x,t,D_x;M_+){\mathbb I}_2 & {\mathbb O}_2  \nonumber \\ 
{\mathbb O}_2 &  {\cal G}(x,t,D_x;M_-){\mathbb I}_2    \\ 
   \end{array}
   \right )[e^{H\cdot }F ] \right.\\
&  &
\left.+i\gamma ^0 \left (
   \begin{array}{cccc}
 {\cal K}_1(x,t,D_x;M_+) {\mathbb I}_2&{\mathbb O}_2  \\
 {\mathbb O}_2 & {\cal K}_1(x,t,D_x;M_-) {\mathbb I}_2 \\
   \end{array}
   \right )[\Phi  (x ) ] \right ]  \,.
 \end{eqnarray}
\medskip

 \section{\bf Proof of the sufficiency part}
 \label{S2}

 \subsection{\bf Some exceptional cases of operator ${\cal K}_1(x,t,D_x;M)$}

Denote $V_\varphi(x,t) $ the solution of the problem 
\begin{eqnarray} 
\label{14}
V_{tt}-  {\mathcal A}(x,\partial_x) V =0, \quad V(x,0)= 0, \quad V_t(x,0)=\varphi (x)\,,
\end{eqnarray}
then $V_\varphi(x,t) $  can related to the solution     $v_{\varphi } $ of (\ref{1.23}) as follows
\[
v_{\varphi }(x,t) =\frac{\partial }{\partial t}V_{\varphi }(x,t)\,.
\]
The kernel       $K_1(z,t;M) $ can be written in the explicit form:
\begin{eqnarray} 
\label{K1}
K_1(r,t;M)
& :=  &
4^{-\frac{M}{H}} e^{M t} \left(\left(1+e^{-H t}\right)^2-(H r)^2\right)^{\frac{M}{H}-\frac{1}{2}} \nonumber \\
&  &
\times  F \left(\frac{1}{2}-\frac{M}{H},\frac{1}{2}-\frac{M}{H};1;\frac{\left(1-e^{-H t} \right)^2-(r H)^2}{\left(1+e^{-H t} \right)^2-(r H)^2}\right) \,.  
\end{eqnarray}
 Further, according to (\ref{K1})   
we obtain 
\begin{eqnarray*}
K_1\left(r,t;- \frac{1}{2}H \right)
& =  &
K_1\left(r,t;  \frac{1}{2}H \right)=
\frac{1}{2}   e^{\frac{1}{2}H t}\,,\\
K_1\left(r,t; \frac{3}{2}H\right)
& =  & 
\frac{1}{4}   e^{-\frac{1}{2}  H t } \left(\left(1-H^2 r^2\right) e^{2 H t}+1\right)\,.
\end{eqnarray*}
Consequently,  by the definition (\ref{K1OPER}) of the operator $ {\cal K}_1$ we write
\begin{eqnarray} 
{\cal K}_1\left(x,t,D_x;\frac{1}{2}H \right)[\varphi   (x ) ]
& =  &
\label{K1minus12} 
  {\cal K}_1\left(x,t,D_x;- \frac{1}{2}H\right)[\varphi   (x ) ] = e^{\frac{1}{2}Ht}  V_{\varphi } (x, \phi (t) )\,,
\end{eqnarray}
and
\begin{eqnarray}
\label{K1plus32}
& &
 {\cal K}_1\left(x,t,D_x;\frac{3}{2}H\right)[\varphi   (x )] \\
& = & 
2\int_{0}^{\phi (t) }  v_{\varphi  } (x,  s)
 \frac{1}{4}  e^{-\frac{1}{2} H t} \left(\left(1-H^2 s^2\right) e^{2 H t}+1\right)    ds \nonumber \\ 
&  =  & 
\frac{1}{2}   e^{ \frac{3}{2}  H t }  \left(1+ e^{-2 H t}\right) V_{\varphi _1 }(x,  \phi (t))   
-H^2 \frac{1}{2}  e^{ \frac{3}{2}  H t }  \phi (t)^2 V_{\varphi  }(x,  \phi (t))
+ H^2     e^{\frac{3}{2}  H t}  \int_{0}^{\phi (t) }    V_{\varphi   } (x,  s) 
  s    ds  \,. \nonumber  
\end{eqnarray}

\subsection{\bf Sufficiency of $m=0,\pm iH$ for the  incomplete   Huygens  principle } 

\label{S4}

 First we  consider the case of $m=0$, then $M_+=M_-=M=\frac{1}{2}H$ and  the kernel $K_1( s,t;M ) $ that is given by (\ref{K1}). 
Consequently, 
for the solution of the generalized Dirac equation with $m=0$ and $F=0$ we obtain from (\ref{solution}) 
 \begin{eqnarray*} 
\Psi (x,t) 
& = &
  e^{-Ht}\left(  \partial_0 {\mathbb I}_4 +  e^{-Ht} \gamma ^k \gamma ^0{\mathcal A}_k(x,\partial_x)- \frac{H}{2}{\mathbb I}_4   \right)  e^{Ht/2} \left (
   \begin{array}{cccc}
   V_{\Phi _0} (x, \phi (t) )   \\
  V_{\Phi _1 } (x, \phi (t) ) \\
   V_{\Phi _ 2} (x, \phi (t) )   \\
   V_{\Phi _3 } (x, \phi (t) ) \\
   \end{array}
   \right ) 
\end{eqnarray*}
 and the Huygens  principle is valid since by the assumption of the theorem it is valid for the problem (\ref{1.23}). 
\medskip

Now we are going to prove  that  $m=\pm iH$  is sufficient for the  incomplete Huygens  principle for the generalized Dirac equation in the de~Sitter spacetime and for the   Huygens  principle for the   Dirac equation in the de~Sitter spacetime. 
For $m=iH$ we have $M_+=  - \frac{1}{2}H  $ and $M_-=  \frac{3}{2}H$.
Then for the operators $ {\cal K}_1\left(x,t,D_x;- \frac{1}{2}H\right)$ and $  {\cal K}_1\left(x,t,D_x;\frac{3}{2}H\right)$
we have representation (\ref{K1minus12}) and (\ref{K1plus32}), respectively. 
Hence the following function 
\[
\Psi (x,t) 
=
  e^{-Ht}\left(  \partial_0 {\mathbb I}_4 +  e^{-Ht} \gamma ^k \gamma ^0{\mathcal A}_k(x,\partial_x)- \frac{H}{2}{\mathbb I}_4 -im \gamma ^0 \right)  \left (
   \begin{array}{cccc}
 {\cal K}_1(x,t,D_x;M_+)[\Phi _0(x ) ]  \\
{\cal K}_1(x,t,D_x;M_+)[ \Phi _1(x )] \\
 {\cal K}_1(x,t,D_x;M_-)[\Phi _2(x )]  \\
 {\cal K}_1(x,t,D_x;M_-)[\Phi _3(x ) ]\\
   \end{array}
   \right ) 
\]
solves the generalized Dirac equation, where the matrices $\gamma ^1\gamma ^0 $, $\gamma ^2\gamma ^0 $,$\gamma ^3\gamma ^0 $ can be written 
as follows
\begin{equation}
\label{5.29} 
\gamma ^1\gamma ^0=
\left(
\begin{array}{cc}
{\mathbb O}_2  & -\sigma _1   \\
-\sigma _1 & {\mathbb O}_2    \\ 
\end{array}
\right), \quad 
 \gamma ^2\gamma ^0=
\left(
\begin{array}{cc}
 {\mathbb O}_2  & -\sigma _2   \\
 -\sigma _2 & {\mathbb O}_2    \\ 
\end{array}
\right),\quad 
  \gamma ^3\gamma ^0=
\left(
\begin{array}{cc}
{\mathbb O}_2  & -\sigma _3   \\
-\sigma _3 & {\mathbb O}_2    \\ 
\end{array}
\right) 
.
\end{equation} 
We split the  initial function  $ \Phi  (x )$ into two parts. First we consider the case of  
\[
\Phi _2(x )=\Phi _3(x )=0\,,
\]
then the solution $\Psi  =\Psi (x,t)  $ is as follows
\begin{eqnarray*} 
\Psi (x,t) 
& = &
 H^2 e^{-Ht}  \left(  \partial_0 {\mathbb I}_4 +  e^{-Ht} \gamma ^k \gamma ^0{\mathcal A}_k(x,\partial_x)- \frac{H}{2}{\mathbb I}_4   -im \gamma ^0 \right) e^{Ht/2}  \left (
   \begin{array}{cccc}
  V_{\Phi _0 } (x, \phi (t) ) \\
 V_{\Phi _1} (x, \phi (t) )\\
0 \\
0\\
   \end{array}
   \right ) \\
& = &
 H^2 e^{-Ht/2}  \left(  \partial_0 {\mathbb I}_4 +  e^{-Ht} \gamma ^k \gamma ^0{\mathcal A}_k(x,\partial_x) +H{\mathbb I}_4 \right)  \left (
   \begin{array}{cccc}
  V_{\Phi _0 } (x, \phi (t) ) \\
 V_{\Phi _1} (x, \phi (t) )\\
0 \\
0\\
   \end{array}
   \right ) \,.
\end{eqnarray*}
Thus, the sufficiency of $m=iH$  for the  incomplete Huygens  principle for the generalized Dirac equation in the de~Sitter spacetime  follows  from the property of solution operator of the problem  (\ref{1.23}). It also proves partial validity of the incomplete   Huygens  principle for the   Dirac equation in the de~Sitter spacetime with respect to  the first  components of the  initial function. 
On the other hand, for the second pair  of  components of the  initial function, that is, for the case of 
\begin{equation}
\label{3.21}
\Phi _0(x )=\Phi _1(x )=0,
\end{equation}
the solution is as follows
\begin{eqnarray*} 
\Psi (x,t) 
& = &
  e^{-Ht}\left[  \left (
   \begin{array}{ccccc}
 1& 0& 0   & 0 \\
   0& 1 &  0 & 0 \\ 
 0 & 0 &  1   & 0 \\
 0& 0& 0   &  1 \\
   \end{array} \right) \partial_0 +  e^{-Ht} \left(
\begin{array}{cccc}
 0 & 0 & 0 & -1 \\
 0 & 0 & -1 & 0 \\
 0 & -1 & 0 & 0 \\
 -1 & 0 & 0 & 0 \\
\end{array}
\right){\mathcal A}_1(x,\partial_x) \right.\\
&  &
+  e^{-Ht} \left(
\begin{array}{cccc}
 0 & 0 & 0 & i \\
 0 & 0 & -i & 0 \\
 0 & i & 0 & 0 \\
 -i & 0 & 0 & 0 \\
\end{array}
\right){\mathcal A}_2(x,\partial_x)
+  e^{-Ht} \left(
\begin{array}{cccc}
 0 & 0 & -1 & 0 \\
 0 & 0 & 0 & 1 \\
 -1 & 0 & 0 & 0 \\
 0 & 1 & 0 & 0 \\
\end{array}
\right){\mathcal A}_3(x,\partial_x)\\
&  &
 \left.- \frac{H}{2} \left(
   \begin{array}{ccccc}
 1& 0& 0   & 0 \\
   0& 1 &  0 & 0 \\ 
 0 & 0 &  1   & 0 \\
 0& 0& 0   &  1 \\
   \end{array} \right) 
-im \left (
   \begin{array}{ccccc}
 1& 0& 0   & 0 \\
   0& 1 &  0 & 0 \\ 
 0 & 0 &  -1   & 0 \\
 0& 0& 0   &  -1 \\
   \end{array} \right)
 \right]  \left (
   \begin{array}{cccc}
0  \\
0\\
 {\cal K}_1(x,t,D_x;M_-)[\Phi _2(x )]  \\
 {\cal K}_1(x,t,D_x;M_-)[\Phi _3(x ) ]\\
   \end{array}
   \right ) \,.
\end{eqnarray*}
For  the components of the function $\Psi (x,t)=(\Psi_0 (x,t),\Psi_1 (x,t),\Psi_2 (x,t),\Psi_3 (x,t))^T $ we obtain
\begin{eqnarray*} 
\Psi_0 (x,t) 
& = &
  e^{-Ht}\Big[   - e^{-Ht} {\mathcal A}_1(x,\partial_x) {\cal K}_1(x,t,D_x;M_-)[\Phi _3(x ) ] \\
&  &
+  e^{-Ht} i{\mathcal A}_2(x,\partial_x) {\cal K}_1(x,t,D_x;M_-)[\Phi _3(x ) ]
-  e^{-Ht}{\mathcal A}_3(x,\partial_x) {\cal K}_1(x,t,D_x;M_-)[\Phi _2(x ) ] \Big], \\
\Psi_1 (x,t) 
& = &
  e^{-Ht}\Big[  -e^{-Ht} {\mathcal A}_1(x,\partial_x) {\cal K}_1(x,t,D_x;M_-)[\Phi _2(x ) ] \\
&  &
-i  e^{-Ht} {\mathcal A}_2(x,\partial_x){\cal K}_1(x,t,D_x;M_-)[\Phi _2(x ) ] 
+  e^{-Ht} {\mathcal A}_3(x,\partial_x){\cal K}_1(x,t,D_x;M_-)[\Phi _3(x ) ] \Big],\\
\Psi_2 (x,t) 
& = &
  e^{-Ht}  \left( \partial_0 -\frac{H}{2} +im \right)  {\cal K}_1(x,t,D_x;M_-)[\Phi _2(x ) ],  \\
\Psi_3 (x,t) 
& = &
  e^{-Ht}    \left( \partial_0 -\frac{H}{2} +im \right) {\cal K}_1(x,t,D_x;M_-)[\Phi _3(x ) ] .
\end{eqnarray*}
It is easy to see that the tails  of the functions $
\Psi_k (x,t)= \left( \partial_0 -\frac{H}{2} +im \right) {\cal K}_1(x,t,D_x;M_-)[\Phi _k(x ) ]$   ($k=2,3$)  
are empty.  Consider now the tails of the terms
\begin{eqnarray}
\label{2.22}
{\mathcal A}_j(x,\partial_x){\cal K}_1(x,t,D_x;M_-)[\Phi _k(x ) ],\quad j= 1,2,3,\quad k=0,1\,.
\end{eqnarray} 
Now we assume that the operator  is the  Dirac operator     (\ref{DE}).    
Then  according to (\ref{K1plus32})   
\begin{eqnarray*} 
{\mathcal A}_j(x,\partial_x){\cal K}_1(x,t,D_x;M_-)[\Phi _k(x ) ] 
& = &
\frac{\partial}{\partial x_j}\Bigg[\frac{1}{2}   e^{ \frac{3}{2}  H t }  \left(1+ e^{-2 H t}\right) V_{\Phi _k }(x,  \phi (t))   \\
&  &
-H^2 \frac{1}{2}  e^{ \frac{3}{2}  H t }  \phi (t)^2 V_{\Phi _k }(x,  \phi (t))
+ H^2     e^{\frac{3}{2}   H t }  \int_{0}^{\phi (t) }    V_{\Phi _k} (x,  s) 
  s    ds    \Bigg] 
\end{eqnarray*}
and the only possible  tail of function (\ref{2.22}) is
\begin{eqnarray*} 
&  &
H^2e^{  \frac{3}{2}Ht}  \frac{\partial}{\partial {x_j}}  \int_{0}^{\phi (t) }   s   
   V _{\Phi _k } (x,  s)   ds  \,.
\end{eqnarray*}
We apply the Kirchhoff's formula and consider, for instance,  the case of $j=3$ in (\ref{2.22}). Then up to unimportant factor, the possible tail is:
\begin{eqnarray*} 
&  & 
 \frac{\partial}{\partial {x_3}}    \int_{0}^{\phi (t) }  s^2   \int \!\! \int_{|y|=1}    \Phi _k  (x+sy) dS_y 
  =   
\frac{\partial}{\partial {x_3}}    \int  \!\!  \int \!\! \int_{|y|\leq \phi (t)} \Phi _k   (x+y) d\,y_1\,dy_2\,dy_3 \\
& =  &
\int  \!\!  \int \!\! \int_{|y|\leq \phi (t)}    \frac{\partial}{\partial {x_3}}  \Phi _k  (x+y) d\,y_1\,dy_2\,dy_3  
  =   
 \int   \!\!  \int \!\! \int_{|y|\leq \phi (t)}    \frac{\partial}{\partial {y_3}} \Phi _k  (x+y) d\,y_1\,dy_2\,dy_3 \\
& =  &
 \int \!\! \int_{y_1^2+y_2^2\leq \phi^2 (t)} dy_1 dy_2 
\int_{-\sqrt{\phi^2 (t)-y_1^2-y_2^2 }}^{ \sqrt{\phi^2 (t)-y_1^2-y_2^2 }}    \frac{\partial}{\partial {y_3}}  \Phi _k  (x_1+y_1, x_2+y_2, x_3+y_3)\, dy_3\\
& =  &
 \int \!\! \int_{y_1^2+y_2^2\leq \phi^2 (t)} \Bigg\{    \Phi _k  \left(x_1+y_1, x_2+y_2, x_3+\sqrt{\phi^2 (t)-y_1^2-y_1^2 }\right)\\
&  &
 \hspace{3cm} -\Phi _k \left(x_1+y_1, x_2+y_2, x_3-\sqrt{\phi^2 (t)-y_1^2-y_1^2 }\right)\Bigg\}dy_1 dy_2 \,.
 \end{eqnarray*}
For every $t>0$ the points
\[
\left(x_1+y_1, x_2+y_2, x_3\pm\sqrt{\phi^2 (t)-y_1^2-y_2^2 }\right) \in {\mathbb R}^3, \quad \mbox{\rm where} \,\, y_1^2+y_1^2\leq \phi^2 (t),
\]
belong to the sphere  of the radius $\phi^2 (t) $ in ${\mathbb R}^3$, that is, the domain of integration does not intersect   the interior of the domain of dependence. Thus, the tail is empty. This completes the proof of the Huygens  principle for the Dirac equation with $m=iH$ in the  de~Sitter spacetime.     
\medskip

For the case of $m=-iH$ we have  $M_+  =  \frac{3}{2}H $,  $M_-=   - \frac{1}{2}H$,  
and the proof is similar to the case of $m=iH $ that was discussed above. Therefore  we skip the proof of the case with $m=-iH$.         \qed 
\medskip

\section{\bf Necessity of $m=0,\pm iH$  for the Huygens  principle. The case of $ m \not= i\frac{H}{2}   +  i\frac{H}{2}\ell, \quad  \ell=0,\pm 1, \pm2,\ldots $} 
\label{S3}

The proof of the necessity parts of the Theorem~\ref{T0.3} is based on the large time asymptotic  of the tail of solution.  The initial data will be chosen radial
having small support containing the origin. The tail   is generated by the functions
\begin{eqnarray*} 
&  &
\left(   \frac{\partial}{\partial t} - \frac{H}{2}  -im\right)  
 K_1\left( r,t; \frac{1}{2}H  + im\right) ,\quad \left(   \frac{\partial}{\partial t} - \frac{H}{2}  +im\right)  
 K_1\left( r,t; \frac{1}{2}H  - im\right)\,.
\end{eqnarray*}
where $K_1( r,t;M ) $~(\ref{K1new}) is the kernel 
of the integral representing   the solution  of the generalized Dirac equation. 
The non-huygensian part of the solution written via these functions contains some integrals over interior of the support  of the initial function. This term simplifies only for the values of  mass mentioned in the theorem and allows us to write solution via solutions of the wave equation without any non-huygensian operation.
\medskip

We set $F=0$ in the representation formula given by Theorem~\ref{T0.3}, then the solution is 
\begin{eqnarray*} 
\Psi (x,t) 
& = &
  e^{-Ht}\left(   \partial_0 {\mathbb I}_4+  e^{-Ht} \gamma ^k \gamma ^0\partial_k- \frac{H}{2}{\mathbb I}_4 -im\gamma ^0\right)   \left (
   \begin{array}{cccc}
 {\cal K}_1(x,t,D_x;M_+)[\Phi _0(x ) ]  \\
{\cal K}_1(x,t,D_x;M_+)[ \Phi _1(x )] \\
 {\cal K}_1(x,t,D_x;M_-)[\Phi _2(x )]  \\
 {\cal K}_1(x,t,D_x;M_-)[\Phi _3(x ) ]\\
   \end{array}
   \right ) \,,
\end{eqnarray*}
where the matrices $\gamma ^1\gamma ^0 $, $\gamma ^2\gamma ^0 $,$\gamma ^3\gamma ^0 $ are written  in (\ref{5.29}).

Consider the solution  of the Cauchy problem with the radial  function $\Phi_0  (x )=\Phi_0  (r ) $,
supp$\,\Phi_0   \subset \left\{x \in {\mathbb R}^{n} ;|x| \leq \min\{ 1/2,  \varepsilon /H \}  \right\}$, $\varepsilon  \in (0,1)$. If we choose the initial data 
\begin{equation} 
\label{3.27}
\Phi (x) = ( \Phi _0(x ) ,0,0,0)^T\,,
\end{equation} 
 then the solution $\Psi  (x,t) = ( \Psi_0  ,\Psi_1 ,\Psi_2 ,\Psi_3 )^T $  is given by
\[
\Psi (x,t) 
  =  
  e^{-Ht}\left(   \partial_0 {\mathbb I}_4+  e^{-Ht} \gamma ^a \gamma ^0\partial_a- \frac{H}{2}{\mathbb I}_4 -im \gamma ^0\right) 
( {\cal K}_1(x,t,D_x;M_+)[\Phi _0(x ) ] ,0,0,0)^T   \,,
\]
while its first component is as follows
\begin{eqnarray*} 
\Psi_0 (x,t) 
& = &
  e^{-Ht}\left(   \partial_0 - \frac{H}{2}  -im\right)   
 {\cal K}_1(x,t,D_x;M_+)[\Phi _0(x ) ] \,.
\end{eqnarray*} 
Now we are going to examine  for large time  the first component of the vector-valued function $\Psi  (x,t)$. We calculate
\begin{eqnarray*} 
\Psi_0 (x,t) 
& = &
  e^{-Ht}\left(  \frac{\partial}{\partial t} - \frac{H}{2}  -im\right)   
 \, 2\int_{0}^{\phi (t) }  v_{\Phi _0 } (x,  s)
  K_1( s,t;M_+)   ds  \,.
\end{eqnarray*}
We can rewrite $\Psi_0 (x,t)$ in the terms of the function $V_{\Phi _0} $ defined in accordance to (\ref{14})
\begin{eqnarray*} 
\Psi_0 (x,t) 
& = &
  e^{-Ht}\left(   \frac{\partial}{\partial t} - \frac{H}{2}  -im\right)   
 \, 2\int_{0}^{\phi (t) } \left( \frac{\partial }{\partial s}V_{\Phi _0}(x,s)  \right)
  K_1( s,t;M_+) \,  ds \,.
\end{eqnarray*}
It follows,
\begin{eqnarray*} 
\Psi_0(x,t) 
& = &
 2   e^{-Ht}\left(   \frac{\partial}{\partial t} - \frac{H}{2}  -im\right)   
    V_{\Phi _0}(x, \phi (t))  K_1(\phi (t),t;M_+)   \\
&   &
 - 2 e^{-2Ht}   
   V_{\Phi _0}(x,\phi (t))   
\left(  \frac{\partial  }{\partial s  } K_1( s,t;M_+) \right)_{s=\phi (t)}  \\
&   &
 - 2 e^{-Ht}  
  \int_{0}^{\phi (t) } V_{\Phi _0}(x,s) \left(   \frac{\partial}{\partial t} - \frac{H}{2}  -im\right)  
 \frac{\partial }{\partial s} K_1( s,t;M_+)   ds \,.
\end{eqnarray*} 
In particular, since $x \in {\mathbb R}^3$, by the Kirchhoff's formula we have  
$
V_{\Phi _0}(0, \tau )=\tau  \Phi _0 \left(\tau \right)\,
$
and
\begin{eqnarray*}
V_{\Phi _0}(0, \phi (t))=\phi (t) \Phi _0 (\phi (t)) =\frac{ 1-e^{-Ht} }{H} \Phi _0 \left(\frac{ 1-e^{-Ht} }{H} \right)=0 
\end{eqnarray*} 
for sufficiently large $t$, that is, if $  1-e^{-Ht}  > H  \varepsilon   $. Consequently, for large $t$ we have 
\begin{eqnarray*} 
\Psi_0 (0,t) 
& = &
 - 2 e^{-Ht}  
  \int_{0}^{\phi (t) } s\Phi _0 ( s) \left(   \frac{\partial}{\partial t} - \frac{H}{2}  -im\right)  
 \frac{\partial }{\partial s} K_1( s,t;M_+)   ds  \nonumber \\ 
& = &
   2 e^{-Ht}  
  \int_{0}^{1 } \left( \frac{\partial }{\partial s} s\Phi _0 ( s) \right) \left(   \frac{\partial}{\partial t} - \frac{H}{2}  -im\right)  
 K_1( s,t;M_+)   ds   \,.
\end{eqnarray*} 
Now we focus on the possible tail of the solution, that is, on the term  generated by the  integral. It is easy to see that
\begin{eqnarray*}
K_1(r,t;M_+)
& =  &
2^{-1-\frac{2 i m}{H}} e^{\frac{1}{2} t (H+2 i m)} \left(\left(e^{-H t}+1\right)^2-H^2 r^2\right)^{\frac{i m}{H}} \\
&  &
\times F \left(-\frac{i m}{H},-\frac{i m}{H};1;\frac{\left(-1+e^{-H t}\right)^2-H^2 r^2}{\left(1+e^{-H t}\right)^2-H^2 r^2}\right)\,.
\end{eqnarray*} 
  
\begin{lemma}
\label{L4.1}
For $m \in {\mathbb C}$ one has
\begin{eqnarray*} 
&  &
\left(   \frac{\partial}{\partial t} - \frac{H}{2}  -im\right)  
 K_1( r,t;M_+) \\
&  = &
2^{-\frac{2 i m}{H}}im e^{\frac{1}{2} t (2 i m-H )} \left(\left(1+e^{-H t}\right)^2-H^2 r^2\right)^{\frac{i m}{H}-2}\\
&  &
\times 
 \Bigg\{  2 \frac{im }{H   }  \left(  1-  e^{-2 H t}- H^2 r^2\right)  F \left(1-\frac{i m}{H},1-\frac{i m}{H};2;\frac{\left( 1-e^{-H t}\right)^2-H^2 r^2}{\left(1+e^{-H t}\right)^2-H^2 r^2}\right)\\
&  &
-  \left(1+e^{-H t}\right) 
 \left(  (1 + e^{-H t})^2-H^2 r^2\right)  F \left(-\frac{i m}{H},-\frac{i m}{H};1;\frac{\left( 1-e^{-H t}\right)^2-H^2 r^2}{\left(1+e^{-H t}\right)^2-H^2 r^2}\right)\Bigg\} \,.
\end{eqnarray*}
\end{lemma}
\medskip

\noindent
{\bf Proof.} We skip the  proof since this statement can be verified by straightforward calculations. 
\hfill $\square$

\begin{proposition}
\label{P4.2}
Assume that $m \in {\mathbb C}$ and 
\[
 m \not= i\frac{H}{2}   +  i\frac{H}{2}\ell, \quad  \ell=0,\pm 1, \pm2,\ldots \,,
\] 
then
\begin{eqnarray*} 
&  &  
2 \frac{im }{H   }  \left(  1-  e^{-2 H t}- H^2 r^2\right)  F \left(1-\frac{i m}{H},1-\frac{i m}{H};2;\frac{\left( 1-e^{-H t}\right)^2-H^2 r^2}{\left(1+e^{-H t}\right)^2-H^2 r^2}\right)\\
&  &
 -  \left(e^{-H t}+1\right) 
 \left(  (1 + e^{-H t})^2-H^2 r^2\right)  F \left(-\frac{i m}{H},-\frac{i m}{H};1;\frac{\left( 1-e^{-H t}\right)^2-H^2 r^2}{\left(1+e^{-H t}\right)^2-H^2 r^2}\right)\\
&  = & 
\dsp 2 \frac{im }{H   }   4 ^{2\frac{i m}{H} }    \frac{  \Gamma(-2\frac{i m}{H} )}{[\Gamma(1-\frac{i m}{H})]^2 }\left(  1-  H^2 r^2\right)^{1-2\frac{i m}{H}}  
      e^{-  2 i  m t} (1  +  R (m,H,r;t))\,,
\end{eqnarray*} 
where  with  large  $T$  the remainder  $R (m,H,r;t) $    can be estimated as follows 
\begin{eqnarray}
\label{4.25}
|R (m,H,r;t)| \leq o(1) \quad   \mbox{\rm for all} \quad  t \geq T \quad  and  \quad   0 \leq r \leq \min \{1/2, 1/(2H)\} \,.
\end{eqnarray}
 \end{proposition}
\medskip

\noindent
{\bf Proof.} Consider the first term. 
In order to simplify notations we denote 
\begin{eqnarray*} 
& &
A:= Hr,\quad \tau := e^{-H t} ,\quad   \\
&  &
z := \frac{\left( 1-e^{-H t}\right)^2-H^2 r^2}{\left(1+e^{-H t}\right)^2-H^2 r^2}  
= \frac{\left( 1-\tau \right)^2-A^2}{\left(1+\tau \right)^2-A^2}=
1-\frac{4 \tau }{1-H^2 r^2 }+\frac{8 \tau ^2}{\left(1-H^2 r^2 \right)^2}+O\left(\tau ^3\right),\\
&  &
1-z 
  =  
\frac{4e^{-H t} }{\left(1+e^{-H t}\right)^2-H^2 r^2} =  
\frac{4\tau  }{\left(1+\tau \right)^2-A^2}= \frac{4 \tau }{1-H^2 r^2 }-\frac{8 \tau ^2}{\left(1-H^2 r^2 \right)^2}+O\left(\tau ^3\right). 
\end{eqnarray*}
There is a 
formula (\ref{15.3.6A}) in Appendix 
that ties together points $z=0$ and $z=1$ of the argument of the hypergeometric function. Hence we have (\ref{15.3.6A}) for all  $1/2<z<1$. 
For all $m \in {\mathbb C}$ such that
\begin{eqnarray*} 
& &
a 
 = b=1-\frac{i m}{H}\,,\quad c=2\,,\quad \frac{i m}{H} \not= 0, \pm 1,\pm 2,\ldots
\quad \Longleftrightarrow \quad m \not=0, \pm iH ,\pm 2 iH,  \pm 3 iH,\ldots\,, 
\end{eqnarray*}
according to  (\ref{15.3.6A}) we can write
\begin{eqnarray*}  
F \left(1-\frac{i m}{H},1-\frac{i m}{H};2;z\right)  
  & \!\!= \!\!& 
\frac{\Gamma(2\frac{i m}{H})}{[\Gamma(1+\frac{i m}{H})]^2 } F\left(1-\frac{i m}{H},1-\frac{i m}{H};1 -2\frac{i m}{H} ; 1-z \right) \\
& &
+\left(1-z \right)^{2\frac{i m}{H} } \frac{ \Gamma(-2\frac{i m}{H} )}{[\Gamma(1-\frac{i m}{H})]^2 } F\left(  1+\frac{i m}{H}   ,  1+\frac{i m}{H}    ;1 +2\frac{i m}{H}  ; 1-z \right) .
\end{eqnarray*} 
From $F \left(a,b;c;0\right) =1$ it follows
\begin{eqnarray*}  
F \left(1-\frac{i m}{H},1-\frac{i m}{H};2;z\right)
  & = &
 \frac{  \Gamma(2\frac{i m}{H})}{[\Gamma(1+\frac{i m}{H})]^2 } \left( 1+ R_0(m,H,r;t) \right) \\
& &
+\left(1-z \right)^{2\frac{i m}{H} } \frac{  \Gamma(-2\frac{i m}{H} )}{[\Gamma(1-\frac{i m}{H})]^2 }  \left( 1+ R_1(m,H,r;t) \right) \,,
\end{eqnarray*} 
where  with  large  $T$  the remainders $R_k(m,H,r;t) $, $k= 0,1$, can be estimated as follows 
\[
|R_k(m,H,r;t)| \leq o(1)  \quad \mbox{\rm for all} \quad  t \geq T \quad  \mbox{\rm and}  \quad   0 \leq r \leq 1/(2H)\,.
\]  
For the second term of the proposition   we note 
\begin{eqnarray} 
\label{3.20}
& &
a 
 = b= -\frac{i m}{H}\,,\quad c=1\,, \quad  c-a-b=1+ 2\frac{i m}{H} \not= 0,   \pm 1, \pm 2,\ldots\,,\\
 \label{3.21b}
&  &
\Longleftrightarrow m \not= i\frac{H}{2}, \,i\frac{H}{2}   \pm i\frac{H}{2} ,\,i\frac{H}{2}   \pm 2i\frac{H}{2},\,i\frac{H}{2}   \pm 3 i\frac{H}{2},\ldots\,, 
\end{eqnarray}
and obtain
\begin{eqnarray*}  
F \left(-\frac{i m}{H},-\frac{i m}{H};1;z\right) 
& = &
\frac{ \Gamma(1+ 2\frac{i m}{H})}{[\Gamma(1+  \frac{i m}{H})]^2} F\left(-\frac{i m}{H}, -\frac{i m}{H} ; - 2\frac{i m}{H} ; 1-z\right) \\
& &
+\left(1-z\right)^{1+ 2\frac{i m}{H}} \frac{\Gamma(-1- 2\frac{i m}{H})}{[\Gamma(- \frac{i m}{H})]^2} F\left(1+  \frac{i m}{H}, 1+  \frac{i m}{H} ; 2+ 2\frac{i m}{H} ; 1-z\right). 
\end{eqnarray*}
Assume that {$\Re(im)\geq 0$}.  
Then   we have 
\begin{eqnarray*} 
F \left(-\frac{i m}{H},-\frac{i m}{H};1;z\right) 
=
  \frac{ \Gamma(1+ 2\frac{i m}{H})}{[\Gamma(1+  \frac{i m}{H})]^2}   
 + o(1)\quad  \mbox{\rm as}  \quad z \nearrow 1 \,.
\end{eqnarray*}
It follows for $\Re(im)\geq 0$ 
\begin{eqnarray*} 
&  &  
2 \frac{im }{H   }  \left(  1-  \tau ^2- H^2 r^2\right)  F \left(1-\frac{i m}{H},1-\frac{i m}{H};2;z\right)\\
&  &
 -  \left(\tau +1\right) 
 \left(  (1 + \tau )^2-H^2 r^2\right)  F \left(-\frac{i m}{H},-\frac{i m}{H};1;z\right)\\
&  = & 
 2 \frac{im }{H   }\left(  1- \tau ^2- H^2 r^2\right) 
\Bigg\{\frac{  \Gamma(2\frac{i m}{H})}{[\Gamma(1+\frac{i m}{H})]^2 } 
+\left(1-z\right)^{2\frac{i m}{H} } \frac{  \Gamma(-2\frac{i m}{H} )}{[\Gamma(1-\frac{i m}{H})]^2 }  + o(1) \Bigg\}\\
&  &
  -  \left(\tau +1\right) 
 \left(  (1 + \tau )^2-H^2 r^2\right)\Bigg\{\frac{ \Gamma(1+ 2\frac{i m}{H})}{[\Gamma(1+  \frac{i m}{H})]^2}   
+ o(1)  \Bigg\}\\
&  = & 
 2 \frac{im }{H   }\left(  1-  \tau ^2- H^2 r^2\right) 
\Bigg\{\frac{  \Gamma(2\frac{i m}{H})}{[\Gamma(1+\frac{i m}{H})]^2 } 
+\left(1-z\right)^{2\frac{i m}{H} } \frac{  \Gamma(-2\frac{i m}{H} )}{[\Gamma(1-\frac{i m}{H})]^2 }  + o(1) \Bigg\}\\
&  &
  -  
 \left(  1  -H^2 r^2\right)\Bigg\{\frac{ \Gamma(1+ 2\frac{i m}{H})}{[\Gamma(1+  \frac{i m}{H})]^2}   
+ o(1)  \Bigg\}\\
&  = & 
 2 \frac{im }{H   }\left(  1-  \tau ^2- H^2 r^2\right) 
\left\{
\left(1-z\right)^{2\frac{i m}{H} } \frac{  \Gamma(-2\frac{i m}{H} )}{[\Gamma(1-\frac{i m}{H})]^2 }  + o(1) \right\}\,.
\end{eqnarray*}
Thus
\begin{eqnarray*}
&  &  
2 \frac{im }{H   }  \left(  1-  e^{-2 H t}- H^2 r^2\right)  F \left(1-\frac{i m}{H},1-\frac{i m}{H};2;z\right)\\
&  &
 -  \left(e^{-H t}+1\right) 
 \left(  (1 + e^{-H t})^2-H^2 r^2\right)  F \left(-\frac{i m}{H},-\frac{i m}{H};1;z\right) \nonumber \\
&  = & 
2 \frac{im }{H   } \left( 4e^{-H t}   \right)^{2\frac{i m}{H} }\left(  1-  H^2 r^2\right)^{1-2\frac{i m}{H}}  
 \frac{  \Gamma(-2\frac{i m}{H} )}{[\Gamma(1-\frac{i m}{H})]^2 }  + o(1) \,, \quad \mbox{\rm as}\quad t \to \infty  \nonumber \,,
\end{eqnarray*}
uniformly with respect to $   r \in (0, \leq \min \{1/2, 1/(2H)\}) $.  
Next we assume that {$  \Re (im)   <  0 $}.
Consider
\begin{eqnarray*}   
F \left(1-\frac{i m}{H},1-\frac{i m}{H};2;z\right)
  & = &
\frac{  \Gamma(2\frac{i m}{H})}{[\Gamma(1+\frac{i m}{H})]^2 } \left( 1+ R_0(m,H,r;t) \right) \\
& &
+\left(1-z \right)^{2\frac{i m}{H} } \frac{  \Gamma(-2\frac{i m}{H} )}{[\Gamma(1-\frac{i m}{H})]^2 }  \left( 1+ R_1(m,H,r;t) \right)\,.
\end{eqnarray*}
Since $  \Re (im)   <  0 $ the principal term of the asymptotics is $\left(1-z \right)^{2\frac{i m}{H} } \frac{  \Gamma(-2\frac{i m}{H} )}{[\Gamma(1-\frac{i m}{H})]^2 }  $ and  
\begin{eqnarray*}
F \left(1-\frac{i m}{H},1-\frac{i m}{H};2;z\right)
  & = &
\left(1-z \right)^{2\frac{i m}{H} } \frac{  \Gamma(-2\frac{i m}{H} )}{[\Gamma(1-\frac{i m}{H})]^2 }  \left( 1+ R_1(m,H,r;t) \right)\,. 
\end{eqnarray*}
Now for parameters $a,b,c$ satisfying (\ref{3.20}),(\ref{3.21b}) 
consider
\begin{eqnarray*}  
F \left(-\frac{i m}{H},-\frac{i m}{H};1;z\right) 
& = &
\frac{ \Gamma(1+ 2\frac{i m}{H})}{[\Gamma(1+  \frac{i m}{H})]^2} F\left(-\frac{i m}{H}, -\frac{i m}{H} ; - 2\frac{i m}{H} ; 1-z\right) \\
& &
+\left(1-z\right)^{1+ 2\frac{i m}{H}} \frac{\Gamma(-1- 2\frac{i m}{H})}{[\Gamma(- \frac{i m}{H})]^2} F\left(1+  \frac{i m}{H}, 1+  \frac{i m}{H} ; 2+ 2\frac{i m}{H} ; 1-z\right). 
\end{eqnarray*}
If $1+ 2\frac{\Re( i m)}{H}=\delta >0 $, then the principal term is $\frac{ \Gamma(1+ 2\frac{i m}{H})}{[\Gamma(1+  \frac{i m}{H})]^2}$ and we obtain 
\begin{eqnarray*}  
F \left(-\frac{i m}{H},-\frac{i m}{H};1;z\right) 
& = &
\frac{ \Gamma(1+ 2\frac{i m}{H})}{[\Gamma(1+  \frac{i m}{H})]^2} +O((1-z)^\delta )\quad \mbox{\rm as}\quad z \nearrow  1. 
\end{eqnarray*}
If $1+ 2\frac{\Re( i m)}{H}=\delta <0 $ then the principal term is $\left(1-z\right)^{1+ 2\frac{i m}{H}} \frac{\Gamma(-1- 2\frac{i m}{H})}{[\Gamma(- \frac{i m}{H})]^2}$ and we obtain 
\begin{eqnarray*}  
F \left(-\frac{i m}{H},-\frac{i m}{H};1;z\right) 
& = &
 \left(1-z\right)^{1+ 2\frac{i m}{H}} \frac{\Gamma(-1- 2\frac{i m}{H})}{[\Gamma(- \frac{i m}{H})]^2}(1 + O((1-z)^{-\delta } )).  
\end{eqnarray*}
If $1+ 2\frac{\Re( i m)}{H}= 0 $ then both terms are equivalent 
and we obtain 
\begin{eqnarray*}  
F \left(-\frac{i m}{H},-\frac{i m}{H};1;z\right) 
& = &
\frac{ \Gamma(1+ 2\frac{i m}{H})}{[\Gamma(1+  \frac{i m}{H})]^2} F\left(-\frac{i m}{H}, -\frac{i m}{H} ; - 2\frac{i m}{H} ; 1-z\right) \\
& &
+\left(1-z\right)^{1+ 2\frac{i m}{H}} \frac{\Gamma(-1- 2\frac{i m}{H})}{[\Gamma(- \frac{i m}{H})]^2} F\left(1+  \frac{i m}{H}, 1+  \frac{i m}{H} ; 2+ 2\frac{i m}{H} ; 1-z\right)\\
& = &
\frac{ \Gamma(1+ 2\frac{i m}{H})}{[\Gamma(1+  \frac{i m}{H})]^2}  
+\left(1-z\right)^{ 2i\frac{\Re( m)}{H}} \frac{\Gamma(-1- 2\frac{i m}{H})}{[\Gamma(- \frac{i m}{H})]^2}  +o(\tau ) . 
\end{eqnarray*}
Thus, if {$1+ 2\frac{\Re( i m)}{H}=\delta >0 $ and $\Re( i m)  <0 \Longleftrightarrow  \Im (m)>0  $ }, then 
\begin{eqnarray*} 
&  &  
2 \frac{im }{H   }  \left(  1-  e^{-2 H t}- H^2 r^2\right)  F \left(1-\frac{i m}{H},1-\frac{i m}{H};2;z\right)\\
&  &
 -  \left(e^{-H t}+1\right) 
 \left(  (1 + e^{-H t})^2-H^2 r^2\right)  F \left(-\frac{i m}{H},-\frac{i m}{H};1;z\right)\\
&  = &  
2 \frac{im }{H   }  \left(  1-  e^{-2 H t}- H^2 r^2\right)  \Bigg[ 
\left(1-z \right)^{2\frac{i m}{H} } \frac{  \Gamma(-2\frac{i m}{H} )}{[\Gamma(1-\frac{i m}{H})]^2 }  \left( 1+ R_1(m,H,r;t) \right)\Bigg]\\
&  &
 -  \left(e^{-H t}+1\right) 
 \left(  (1 + e^{-H t})^2-H^2 r^2\right) \Bigg[ \frac{ \Gamma(1+ 2\frac{i m}{H})}{[\Gamma(1+  \frac{i m}{H})]^2} +O((1-z)^\delta )\Bigg] \\
&  = &  
2 \frac{im }{H   }  \left(  1-   H^2 r^2\right)  \Bigg[ 
\left(1-z \right)^{2\frac{i m}{H} } \frac{  \Gamma(-2\frac{i m}{H} )}{[\Gamma(1-\frac{i m}{H})]^2 }  \left( 1+ R_1(m,H,r;t) \right)\Bigg]\\
&  &
 -    
 \left( 1  -H^2 r^2\right) \Bigg[ \frac{ \Gamma(1+ 2\frac{i m}{H})}{[\Gamma(1+  \frac{i m}{H})]^2} +O((1-z)^\delta )\Bigg]\\
&  = &  
2 \frac{im }{H   }  \left(  1-   H^2 r^2\right)   
\left(1-z \right)^{-2\frac{\Im( m)}{H} } \left(1-z \right)^{2\frac{i\Re( m)}{H} }\frac{  \Gamma(-2\frac{i m}{H} )}{[\Gamma(1-\frac{i m}{H})]^2 }  \left( 1+ o(1) \right) \\
&  = &  
2 \frac{im }{H   }  \left(  1-   H^2 r^2\right)   
\left(\frac{4\tau  }{\left(1+\tau \right)^2-A^2} \right)^{-2\frac{\Im( m)}{H} + 2\frac{i\Re( m)}{H} } \frac{  \Gamma(-2\frac{i m}{H} )}{[\Gamma(1-\frac{i m}{H})]^2 }  \left( 1+ o(1) \right)\\
&  = &    
4^{2\frac{im }{H   }}2 \frac{im }{H   }  \left(  1-   A^2\right)^{1-2\frac{im }{H   }}   
   \tau   ^{ 2\frac{ i m }{H} }  \frac{  \Gamma(-2\frac{i m}{H} )}{[\Gamma(1-\frac{i m}{H})]^2 }  \left( 1+ o(1) \right)\,, \quad \mbox{\rm as}\quad t \to \infty\,.
\end{eqnarray*}
Thus, if {$1+ 2\frac{\Re( i m)}{H}=\delta <0 $ and $\Re( i m)  <0 \Longleftrightarrow  \Im (m)>0  $ }, then 
\begin{eqnarray*} 
&  &  
2 \frac{im }{H   }  \left(  1-  e^{-2 H t}- H^2 r^2\right)  F \left(1-\frac{i m}{H},1-\frac{i m}{H};2;z\right)\\
&  &
 -  \left(e^{-H t}+1\right) 
 \left(  (1 + e^{-H t})^2-H^2 r^2\right)  F \left(-\frac{i m}{H},-\frac{i m}{H};1;z\right)\\
&  = & 
2 \frac{im }{H   }  \left(  1-  e^{-2 H t}- H^2 r^2\right) \Bigg[\left(1-z \right)^{2\frac{i m}{H} } \frac{  \Gamma(-2\frac{i m}{H} )}{[\Gamma(1-\frac{i m}{H})]^2 }  \left( 1+ R_1(m,H,r;t) \right) \Bigg] \\
&  &
 -  \left(e^{-H t}+1\right) 
 \left(  (1 + e^{-H t})^2-H^2 r^2\right) \Bigg[ \left(1-z\right)^{1+ 2\frac{i m}{H}} \frac{\Gamma(-1- 2\frac{i m}{H})}{[\Gamma(- \frac{i m}{H})]^2}(1 + O((1-z) )) \Bigg]\\
&  = & 
2 \frac{im }{H   }  \left(  1-    H^2 r^2\right)  \left(1-z \right)^{2\frac{i m}{H} } \frac{  \Gamma(-2\frac{i m}{H} )}{[\Gamma(1-\frac{i m}{H})]^2 }  \left( 1+  O((1-z) ) \right) \\
&  = & 
 4   ^{2\frac{i m}{H} } 2 \frac{im }{H   }  \left(  1-    H^2 r^2\right)^{1-2\frac{i m}{H}}   \tau   ^{ 2\frac{im }{H} }   \frac{  \Gamma(-2\frac{i m}{H} )}{[\Gamma(1-\frac{i m}{H})]^2 }  \left( 1+  O((1-z) ) \right)  \,.
\end{eqnarray*}
Thus, if {$1+ 2\frac{\Re( i m)}{H}= 0 $ and $\Re( i m)=-\frac{H}{2}  <0 \Longleftrightarrow  \Im (m)=\frac{H}{2}>0  $ }, then 
\begin{eqnarray*} 
&  &  
2 \frac{im }{H   }  \left(  1-  e^{-2 H t}- H^2 r^2\right)  F \left(1-\frac{i m}{H},1-\frac{i m}{H};2;z\right)\\
&  &
 -  \left(e^{-H t}+1\right) 
 \left(  (1 + e^{-H t})^2-H^2 r^2\right)  F \left(-\frac{i m}{H},-\frac{i m}{H};1;z\right)\\
&  = & 
2 \frac{im }{H   }  \left(  1-  e^{-2 H t}- H^2 r^2\right)\Bigg[\left(1-z \right)^{2\frac{i m}{H} } \frac{  \Gamma(-2\frac{i m}{H} )}{[\Gamma(1-\frac{i m}{H})]^2 }  \left( 1+ R_1(m,H,r;t) \right)\Bigg]   \\
&  &
 -  \left(e^{-H t}+1\right) 
 \left(  (1 + e^{-H t})^2-H^2 r^2\right)\Bigg[\frac{ \Gamma(1+ 2\frac{i m}{H})}{[\Gamma(1+  \frac{i m}{H})]^2}  
+\left(1-z\right)^{ 2i\frac{\Re( m)}{H}} \frac{\Gamma(-1- 2\frac{i m}{H})}{[\Gamma(- \frac{i m}{H})]^2}  +o(\tau )  \Bigg]   \\
&  = & 
2 \frac{im }{H   }  \left(  1-  e^{-2 H t}- H^2 r^2\right)
\Bigg[\left(1-z \right)^{-1+2i\frac{\Im(i m)}{H} } \frac{  \Gamma(-2\frac{i m}{H} )}{[\Gamma(1-\frac{i m}{H})]^2 }  \left( 1+ R_1(m,H,r;t) \right)\Bigg]   \\
&  &
 -  \left(e^{-H t}+1\right) 
 \left(  (1 + e^{-H t})^2-H^2 r^2\right)\Bigg[\frac{ \Gamma(1+ 2\frac{i m}{H})}{[\Gamma(1+  \frac{i m}{H})]^2}  
+\left(1-z\right)^{ 2i\frac{\Re( m)}{H}} \frac{\Gamma(-1- 2\frac{i m}{H})}{[\Gamma(- \frac{i m}{H})]^2}  +o(\tau )  \Bigg]   \\
&  = & 
2 \frac{im }{H   }  \left(  1-  H^2 r^2\right) \left(1-z \right)^{2\frac{i m}{H} } \frac{  \Gamma(-2\frac{i m}{H} )}{[\Gamma(1-\frac{i m}{H})]^2 }  \left( 1+ o(\tau )) \right)\\
&  = & 
 4   ^{2\frac{i m}{H} } 2 \frac{im }{H   }  \left(  1-    H^2 r^2\right)^{1-2\frac{i m}{H}}   \tau   ^{2\frac{i m}{H} }   \frac{  \Gamma(-2\frac{i m}{H} )}{[\Gamma(1-\frac{i m}{H})]^2 }  \left( 1+  O((1-z) ) \right)\,.
\end{eqnarray*}
 The proposition is proved. \qed
\medskip

\noindent
{\bf The necessity of  $ m \not= i\frac{H}{2}   +  i\frac{H}{2}\ell$, $\ell=0,\pm 1, \pm2,\ldots $ in  Theorem~\ref{T0.3}.}
 Thus, from Lemma~\ref{L4.1} and Proposition~\ref{P4.2} we have
\begin{eqnarray*}
\Psi_0 (0,t) 
& = &
   2 e^{-Ht}  
  \int_{0}^{1 } \left( \frac{\partial }{\partial r} r\Phi _0 ( r) \right) \Bigg[ 2^{-\frac{2 i m}{H}}im e^{\frac{1}{2} t (2 i m-H )} \left(\left(1+e^{-H t}\right)^2-H^2 r^2\right)^{\frac{i m}{H}-2}\\
&  &
\times 
 \Bigg\{  2 \frac{im }{H   }  \left(  1-  e^{-2 H t}- H^2 r^2\right)  F \left(1-\frac{i m}{H},1-\frac{i m}{H};2;\frac{\left( 1-e^{-H t}\right)^2-H^2 r^2}{\left(1+e^{-H t}\right)^2-H^2 r^2}\right)\\
&  &
-  \left(e^{-H t}+1\right) 
 \left(  (1 + e^{-H t})^2-H^2 r^2\right)  F \left(-\frac{i m}{H},-\frac{i m}{H};1;\frac{\left( 1-e^{-H t}\right)^2-H^2 r^2}{\left(1+e^{-H t}\right)^2-H^2 r^2}\right)\Bigg\} \Bigg] dr \\
& = &
   2 e^{-Ht}  
  \int_{0}^{1 } \left( \frac{\partial }{\partial r} s\Phi _0 ( r) \right) \Bigg[ 2^{-\frac{2 i m}{H}}im e^{\frac{1}{2} t (2 i m-H )} \left(\left(1+e^{-H t}\right)^2-H^2 r^2\right)^{\frac{i m}{H}-2}\\
&  &
\times 
 \Bigg\{ 2 \frac{im }{H   }   4 ^{2\frac{i m}{H} }   e^{-2 i m   t}   \frac{  \Gamma(-2\frac{i m}{H} )}{[\Gamma(1-\frac{i m}{H})]^2 }\left(  1-  H^2 r^2\right)^{1-2\frac{i m}{H}}  
   +  R (m,H,r;t)  \Bigg\} \Bigg] dr   \\
& = &
   2 e^{-Ht}  
  \int_{0}^{1 } \left( \frac{\partial }{\partial r} r\Phi _0 ( r) \right) \Bigg[ 2^{-\frac{2 i m}{H}}im e^{\frac{1}{2} t (2 i m-H )} \left(\left(1+e^{-H t}\right)^2-H^2 r^2\right)^{\frac{i m}{H}-2}\\
&  &
\times 
 2 \frac{im }{H   }   4 ^{2\frac{i m}{H} }   e^{-2 i m   t}   \frac{  \Gamma(-2\frac{i m}{H} )}{[\Gamma(1-\frac{i m}{H})]^2 }\left(  1-  H^2 r^2\right)^{1-2\frac{i m}{H}}  
   +  R (m,H,r;t)  \Bigg] dr  \\
& = &
   -  e^{\frac{1}{2} t (-2 i m-3H )}H^{-1}  m^2  
      2 ^{2\frac{i m}{H} +2}  \frac{  \Gamma(-2\frac{i m}{H} )}{[\Gamma(1-\frac{i m}{H})]^2 }
 \\
&  &
\times    \int_{0}^{1 }  \Bigg[ \left( \frac{\partial }{\partial r}r\Phi _0 ( r) \right)   \left(  1-  H^2 r^2\right)^{-1- \frac{i m}{H}}  
   +   R (m,H,r;t)  \Bigg] dr \,.
\end{eqnarray*} 
Next we choose $ \Phi _0 $ such that  
 \begin{eqnarray*}
 \int_{0}^{1 } \left( \frac{\partial }{\partial r}r\Phi _0 ( r) \right) \left(  1-  H^2 r^2\right)^{-1- \frac{i m}{H}}\,dr \not= 0\,.
\end{eqnarray*}  
The last equation shows that for $  m \not= i\frac{H}{2}   +  i\frac{H}{2}\ell, \quad  \ell=0,\pm 1, \pm2,\ldots $, the value  $\Psi_1 (0,t)$ for large time depends on the values of the initial function inside of the characteristic conoid. This completes the proof of necessity of such values of $m$. 
 
\section {{\bf Necessity of $m=0,\pm iH$  for the Huygens  principle. Case of $ m = i\frac{H}{2}   +  i\frac{H}{2}\ell, \quad  \ell=0,\pm 1, \pm2,\ldots  $, and    $m\not=0\Longleftrightarrow \ell \not=-1 $}} 

We remind that after Section~\ref{S3} to complete the proof it  remains  to consider the following values of mass: $m= i\frac{H}{2}   +  i\frac{H}{2}\ell, \quad  \ell=0,\pm 1, \pm2,\ldots  $, and $m\not=0\Longleftrightarrow \ell \not=-1 $.  
For all these values we have
\begin{eqnarray*} 
\left(   \frac{\partial}{\partial t} - \frac{H}{2}  -im\right)  
 K_1( r,t;M_+)  
& = &
\left(   \frac{\partial}{\partial t}    +  \frac{H}{2}\ell\right)  
 K_1\left( r,t;-\frac{H}{2}\ell\right) \\
&  = &
H 2^\ell (1+\ell )  e^{-\frac{1}{2} H (\ell+2) t} \left(\left(1+e^{-H t}\right)^2-H^2 r^2\right)^{-\frac{1}{2} (\ell+5)}\\
&  &
\times  {\mathcal F} \left ( e^{-H t},H  r ;\ell;\frac{\left(1-e^{-H t}\right)^2-H^2 r^2}{\left(1+e^{-H t}\right)^2-H^2 r^2} \right) \,,
\end{eqnarray*}
where the function ${\mathcal F} $ is defined as follows 
\begin{eqnarray*}
&  &
{\mathcal F} \left ( e^{-H t},H  r ;\ell;\frac{\left(1-e^{-H t}\right)^2-H^2 r^2}{\left(1+e^{-H t}\right)^2-H^2 r^2} \right)\\
& := &
\left(1+e^{-H t} \right) \left(\left(1+e^{-H t} \right)^2-H^2 r^2\right)  
F \left(\frac{\ell+1}{2},\frac{\ell+1}{2};1;\frac{\left(1-e^{-H t}\right)^2-H^2 r^2}{\left(1+e^{-H t}\right)^2-H^2 r^2}\right)\\
&  &
+(\ell+1) \left(1-e^{-2 H t}-H^2 r^2 \right)  
F \left(\frac{\ell+3}{2},\frac{\ell+3}{2};2;\frac{\left(1-e^{-H t}\right)^2-H^2 r^2}{\left(1+e^{-H t}\right)^2-H^2 r^2}\right)\,.
\end{eqnarray*}
We are going to study a large time asymptotics  of ${\mathcal F}  $. We cannot apply  (\ref{15.3.6A}) since each term of the  formula (\ref{15.3.6A}) has a pole when $c=a+b\pm k$, ($k=0,1,2,\ldots$). 
We split all possible $m = i\frac{H}{2}   +  i\frac{H}{2}\ell, \quad  \ell=0,\pm 1, \pm2,\ldots  $, ($m\not=0\Longleftrightarrow \ell \not=-1 $) into next seven sets:
\begin{enumerate}
\item $\ell=2k+1 $, $k=  -2,-3, -4,\ldots $\,,

\item $\ell=2k+1 $, $ k =1,2,3,\ldots$\,,

\item $\ell=2k $, $k=1,2,3,\ldots$\,,
\item $\ell=2k $, $k=-2,-3,-4,\ldots$\,,
\item $ m =  -  i\frac{H}{2}  $,\, that is
 {$\ell=2k $, $k=-1 $}, $\ell =-2 $\,,
\item $ m = i\frac{H}{2} $,  \, that is
 {$\ell=2k $, $k=\ell =0 $}\,,
\item $ m = iH $, $\ell= 1 $,  {$\ell=2k+1 $, $ k =0$}.
\end{enumerate}
\medskip

\section{The case of odd $\ell=2k+1 $, $k=0,\pm 1, \pm2,\ldots $}

In order to complete the proof, it remains  to consider the following values of mass:
\[
 m = i\frac{H}{2}   +  i\frac{H}{2}\ell, \quad  \ell=0,\pm 1, \pm2,\ldots ,\quad \ell \not=-1\,.
\] 
In this section consider the case of $\ell=2k+1 $, $k=0,\pm 1, \pm2,\ldots $, then
\begin{eqnarray*} 
& &
{\mathcal F} \left ( e^{-H t},H  r ;\ell;\frac{\left(1-e^{-H t}\right)^2-H^2 r^2}{\left(1+e^{-H t}\right)^2-H^2 r^2} \right)\\
& = &  
\left(1+e^{-H t} \right) \left(\left(1+e^{-H t} \right)^2-H^2 r^2\right)  
F \left(k+1,k+1;1;\frac{\left(1-e^{-H t}\right)^2-H^2 r^2}{\left(1+e^{-H t}\right)^2-H^2 r^2}\right)\\
&  &
+(2k+2) \left(1-e^{-2 H t}-H^2 r^2 \right)  
F \left(k+2,k+2;2;\frac{\left(1-e^{-H t}\right)^2-H^2 r^2}{\left(1+e^{-H t}\right)^2-H^2 r^2}\right).
\end{eqnarray*}

\subsection{The case of negative odd $\ell=2k+1 $, $k=  -2,-3, \ldots $}

Denote $A:= H r  \in [0,1/2]$, $\tau =e^{-H t} $, and $z :=  \frac{\left( 1-\tau \right)^2-A^2}{\left(1+\tau \right)^2-A^2}$.  
Since 
\begin{eqnarray*} 
F (-n,-n;1;z)
& = &
1+\sum _{j=1}^n \frac{ \Gamma (n+1)^2}{\Gamma (j+1)^2 \Gamma (n-j +1)^2} z^j\,,\\
F (-n,-n;2;z)
& = &
1+\sum _{j=1}^n \frac{ \Gamma (n+1)^2}{\Gamma (j+2) \Gamma (j+1) \Gamma (n-j+1)^2}z^j \,,
\end{eqnarray*}
we have with $-n=k+1$ and $-n=k+2$ 
\begin{eqnarray*} 
& &
{\mathcal F} \left ( e^{-H t},H  r ;\ell;\frac{\left(1-e^{-H t}\right)^2-H^2 r^2}{\left(1+e^{-H t}\right)^2-H^2 r^2} \right)\\
& = & 
\left(1+\tau  \right) \left(\left(1+\tau  \right)^2-A^2\right)  
F \left(k+1,k+1;1;z\right) 
+(2k+2) \left(1-\tau ^2-A^2 \right)  
F \left(k+2,k+2;2;z\right)\\
& = &    
\left(1+\tau  \right) \left(\left(1+\tau  \right)^2-A^2\right)  
\Bigg\{1+\sum _{j=1}^{- k-1 } \frac{ \Gamma (-k)^2}{\Gamma (j+1)^2 \Gamma (- k -j  )^2}  \Bigg[\frac{\left(1-\tau \right)^2-A^2}{\left(1+\tau \right)^2-A^2}\Bigg] ^j\Bigg\}\\
&  &
+(2k+2) \left(1-\tau ^2-A^2 \right)  
\Bigg\{1+\sum _{j=1}^{- k-2} \frac{ \Gamma (- k-1)^2}{\Gamma (j+2) \Gamma (j+1) \Gamma (- k -j-1)^2} \Bigg[\frac{\left(1-\tau \right)^2-A^2}{\left(1+\tau \right)^2-A^2}\Bigg] ^l\Bigg\}.
\end{eqnarray*}
For the function
$
{\mathcal F} \left ( \tau ,A ;\ell;\frac{ (1-\tau  )^2-A^2}{ (1+\tau  )^2-A^2} \right)
$
with these values of $\ell $ at $\tau =0 $ we have 
\begin{eqnarray*}  
{\mathcal F} \left ( 0 ,A ;\ell;1 \right) 
& = & 0.
\end{eqnarray*} 
Then
\begin{eqnarray*}
\partial_\tau {\mathcal F} \left ( \tau ,A ;\ell;\frac{ (1-\tau  )^2-A^2}{ (1+\tau  )^2-A^2} \right)
& = & 
\left(3 (\tau +1)^2-A^2\right) F \left(k+1,k+1;1;\frac{ (1-\tau  )^2-A^2}{ (1+\tau  )^2-A^2}\right)\\
&  &
+\frac{4(k+1)}{\tau ^4-2 \left(A^2+1\right) \tau ^2+\left(A^2-1\right)^2}\Bigg[  (A-\tau +1) (A+\tau -1)\\
&  &
\times   \left(A^2+2 \tau ^3+3 \tau ^2-1\right)   F\left(k+2,k+2;2;\frac{(A-\tau +1) (A+\tau -1)}{(A-\tau -1) (A+\tau +1)}\right)\\
&  &
\left.-2 (k+2) \left(A^2+\tau ^2-1\right)^2 F \left(k+2,k+3;2;\frac{ (1-\tau  )^2-A^2}{ (1+\tau  )^2-A^2}\right)\right)\Bigg]
\end{eqnarray*}
and
\begin{eqnarray*} 
\partial_\tau {\mathcal F} \left ( \tau ,A ;\ell;\frac{ (1-\tau  )^2-A^2}{ (1+\tau  )^2-A^2} \right)\Bigg|_{\tau =0}
& = &
 -\frac{(-2  k-2))! }{(2 k+3)   [(-k-1)!]^2}\left(A^2 (2 k+3)+1\right)\,, \quad k=-3,-4,\ldots\,.
\end{eqnarray*}
If $k=-2$, then $\ell=-3$  and  the function    of Lemma~\ref{L4.1}  is simplified to the function  independent of $r$ that after substitution in $\Psi _0(0,t) $ creates a   huygensian  part 
and cannot be used to get a contradiction. In fact for $\ell=-3$ the  mass is $m=-iH$, and   the equation is huygensian. 
This explains why we exclude $\ell=-3 $ from the further consideration.

\begin{lemma} 
For $n=-1,-2,\ldots  $ the polynomials $F (n+1,n+1;2;1-x) $ and $F (n,n;1;1-x) $ as $x\searrow 0 $ satisfy
\begin{eqnarray*} 
&  &
 F (n+1,n+1;2;1-x )   = \frac{\Gamma (-2 n)}{\Gamma (1-n)^2}+ x\frac{1}{2} (n+1)^2\frac{2\Gamma ( -2 n-1)}{\Gamma (1-n )^2} +x^2 O(1),\quad n=-1,-2,\ldots \,,\\
&  &
 F (n,n;1;1-x )  = \frac{\Gamma (1-2 n)}{\Gamma (1-n)^2}+ xn^2\frac{ \Gamma ( -2 n )}{\Gamma (1-n )^2}+ x^2 O(1),\quad n=-1,-2,\ldots \,,\\
&  &
 2 n F (n+1,n+1;2;1 )+ F (n,n;1;1 )  = 0,\quad n=-1,-2,\ldots \,,\\
&  &
\frac{d}{d x}   F (n+1,n+1;2;1-x) \Big|_{x=0} =-\frac{1}{2} (n+1)^2\frac{2\Gamma ( -2 n-1)}{\Gamma (1-n )^2}\,,\\
&  &
\frac{d}{d x}    F (n,n;1;1-x) \Big|_{x=0}=-n^2\frac{ \Gamma ( -2 n )}{\Gamma (1-n )^2}\,,\\
&  &
\frac{d}{d x} \left(2 n F (n+1,n+1;2;1-x)+ F (n,n;1;1-x)\right)\Big|_{x=0}\\
& &
 = -\frac{n}{\Gamma (1-n)^2}  (  3 n  +2 )\Gamma (-2 n-1)=A(n)<0,\quad n=-1,-2,\ldots \,.
\end{eqnarray*}
Hence,
\begin{eqnarray*} 
&  &
 2 n F (n+1,n+1;2;1-x )+ F (n,n;1;1-x )  \\
& = &
x \frac{n}{\Gamma (1-n)^2}  (  3 n  +2 )\Gamma (-2 n-1)+x^2O(1),\quad x \searrow  0, \quad n=-1,-2,\ldots \,.
\end{eqnarray*}

\end{lemma}
\medskip

\noindent
{\bf Proof.} For $n =-1,-2,\ldots$ we have
$ c-a-b =-2n>0$ and the hypergeometric functions are polynomials. We apply (\ref{15.3.6A}) and obtain
\[ 
 F (n+1,n+1;2;1 )  
 = 
\frac{\Gamma (-2 n)}{\Gamma (1-n)^2}  \quad \mbox{\rm if} \quad n<0\,, \quad\mbox{\rm and}\quad
F (n,n;1;1 )
  =  
 \frac{\Gamma (1-2 n)}{\Gamma (1-n)^2}  \quad \mbox{\rm if} \quad n\leq 0\,,
\]
then
\begin{eqnarray*} 
2 n F (n+1,n+1;2;1 )+ F (n,n;1;1 )
& = & 
2 n \frac{\Gamma (-2 n)}{\Gamma (1-n)^2}+ \frac{\Gamma (1-2 n)}{\Gamma (1-n)^2}=0 \quad \mbox{\rm if} \quad  n<0\,.
\end{eqnarray*}
Further according to \cite[(7) Ch 2]{B-E} and (\ref{15.3.6A}) we obtain
\begin{eqnarray*} 
\frac{d}{d x}  F (n+1,n+1;2;1-x) 
& = & 
-\frac{1}{2} (n+1)^2 F (n+2,n+2;3;1-x)\,,\\
\frac{d}{d x}  F (n,n;1;1-x)
& = &  
-n^2F(n+1,n+1;2;1-x)
\end{eqnarray*}
and
\begin{eqnarray*} 
\frac{d}{d x}  F (n+1,n+1;2;1-x) \Big|_{x=0}
& = & 
-\frac{1}{2} (n+1)^2\frac{2\Gamma ( -2 n-1)}{\Gamma (1-n )^2}\,,\quad n=-1,-2,\ldots\,,\\
\frac{d}{d x}  F (n,n;1;1-x)\Big|_{x=0}
& = &  
-n^2\frac{ \Gamma ( -2 n )}{\Gamma (1-n )^2},\quad n=-1,-2,\ldots\,.
\end{eqnarray*}
Then for $n=-1,-2,-3,\ldots$
\begin{eqnarray*} 
&  &
\frac{d}{d x} \left(2 n F (n+1,n+1;2;1-x)+ F (n,n;1;1-x)\right)\Big|_{x=0} \\
& = &
\left( -2 n\frac{1}{2} (n+1)^2 F (n+2,n+2;3;1-x)-n^2F(n+1,n+1;2;1-x)\right)\Big|_{x=0}\,.
\end{eqnarray*}
For $3-2(n+2)  >0$ and $2-2(n+1) >0 $ (that is, $n=-1,-2,\ldots $) we obtain 
\[ 
\frac{d}{d x} \left(2 n F (n+1,n+1;2;1-x)+ F (n,n;1;1-x)\right)\Big|_{x=0}  
  =  
 -\frac{n}{\Gamma (1-n)^2}  (  3 n  +2 )\Gamma (-2 n-1) \,.
\]
Lemma is proved. \hfill  $\square$ 

 \begin{corollary} 
 \label{C5.2}
 For every $k=- 2, -3,\ldots $  :
\begin{eqnarray*} 
&  &  
\left(1+\tau  \right) \left(\left(1+\tau  \right)^2-A^2 \right)  
F \left(k+1,k+1;1;z\right)
+(2k+2) \left(1-\tau ^{ 2 }-A^2 \right)  
F \left(k+2,k+2;2;z\right)\\
&  = & 
 2 \frac{(k+1)}{\Gamma (-k)^2}  \left[ (A^2 (- 2 k-3 )+12 k+19 ) \Gamma (-2 k-3)+4\Gamma (-k)^2\right]\tau 
+\tau ^2 O(1),
\end{eqnarray*}
as $\tau \searrow 0 $. Here for every $k= - 2 , -3,\ldots $ if $A $ is such that 
\begin{equation}
\label{A}
A<  A(k) =\min \left\{ \frac{1}{2}, \left(  \frac{1}{(2 k+3)}\left[ 12 k+19   +4\frac{\Gamma (-k)^2}{\Gamma (-2 k-3)}\right]\right)^{1/2} \right\},  
\end{equation}
then   
\[
  (k+1)  \left(\left[-A^2  (2 k+3) +12 k+19\right] \frac{\Gamma (-2 k-3)}{\Gamma (-k)^2}+4\right) >0\,.  
\]
\end{corollary}
\medskip

\noindent
{\bf Proof.} 
First of all, we note that inequality 
\[
4\frac{\Gamma (-k)^2}{\Gamma (-2 k-3)}\leq 2 
\]
implies
\[
 \left[-A^2  (2 k+3) +12 k+19\right] \frac{\Gamma (-2 k-3)}{\Gamma (-k)^2}+4  <0,  
\]
for all $A$ satisfying (\ref{A}).  
Next we consider
\begin{eqnarray*} 
&  &  
\left(1+\tau  \right) \left(\left(1+\tau  \right)^2-A^2 \right)  
F \left(k+1,k+1;1;z\right)\\
&  = & 
\Big(\left(1-A^2\right)+\left(3-A^2\right) \tau +3 \tau ^2+\tau ^3 \Big) \Bigg[ \frac{\Gamma (1-2(k+1))}{\Gamma (1-(k+1))^2}\\
&  &
 + \frac{4\tau  }{\left(1+\tau \right)^2-A^2}(k+1)^2\frac{ \Gamma ( -2(k+1) )}{\Gamma (1-(k+1) )^2}
+ \left(\frac{4\tau  }{\left(1+\tau \right)^2-A^2}\right)^2 O(1)\Bigg]\,.
\end{eqnarray*}
We can continue it as follows
\begin{eqnarray*}
&  = & 
\Big(\left(1-A^2\right)+\left(3-A^2\right) \tau +3 \tau ^2+\tau ^3 \Big)\\
&  &
\times \Bigg[ \frac{\Gamma ( -2 k-1 )}{\Gamma ( -k)^2}+ \left(-\frac{4 \tau }{A^2-1}-\frac{8 \tau ^2}{\left(A^2-1\right)^2}-\frac{4 \left(A^2+3\right) \tau ^3}{\left(A^2-1\right)^3}+O\left(\tau ^4\right)\right)(k+1)^2\frac{ \Gamma ( -2 k-2 )}{\Gamma ( - k  )^2}\\
&  &
+ \left(\frac{16 \tau ^2}{\left(A^2-1\right)^2}+\frac{64 \tau ^3}{\left(A^2-1\right)^3}+O\left(\tau ^4\right)\right) \Bigg]\\
\end{eqnarray*}
and, consequently,  
\begin{eqnarray*} 
&  &  
\left(1+\tau  \right) \left(\left(1+\tau  \right)^2-A^2 \right)  
F \left(k+1,k+1;1;z\right)\\
&  = & 
-\frac{\left(A^2-1\right) \Gamma (-2 k-1)}{\Gamma (-k)^2} 
+\tau \left( -\frac{\left(A^2+2 k-1\right) \Gamma (-2 k-1)}{\Gamma (-k)^2}\right) \\
&  &
+\tau ^2\left( \frac{(1-2 k) \Gamma (-2 k-1)}{\Gamma (-k)^2}-\frac{16}{A^2-1}\right)  
+\tau ^3\left(\frac{\Gamma (-2 k-1)}{\Gamma (-k)^2}-\frac{16 \left(A^2+1\right)}{\left(A^2-1\right)^2}\right)+O\left(\tau ^4\right)\,.
\end{eqnarray*}
For the term with $ F \left(k+2,k+2;2;z\right)$ we obtain 
\begin{eqnarray*}
&  &
(2k+2) \left(1-\tau ^{ 2 }-A^2 \right)  
F \left(k+2,k+2;2;z\right)\\
&  = & 
(2k+2) \left(1-\tau ^{ 2 }-A^2 \right)  
\Bigg[\frac{\Gamma (-2k-2)}{\Gamma (-k)^2}
+ \left(1-z\right)\frac{1}{2} (k+2)^2\frac{2\Gamma ( -2 k-3)}{\Gamma (-k )^2} +\left(1-z\right)^2 O(1) \Bigg] \\ 
& = &
-\frac{2 \left(\left(A^2-1\right) (k+1) \Gamma (-2 k-2)\right)}{\Gamma (-k)^2}
+ \tau \left( 8 (k+1) \left(\frac{(k+2)^2 \Gamma (-2 k-3)}{\Gamma (-k)^2}+1\right)\right)
\\
&  &
+ \tau ^2 \frac{2 (k+1)}{A^2-1} \left(\frac{\left(A^2 (2 k+3)+8 k^2+30 k+29\right) \Gamma (-2 k-3)}{\Gamma (-k)^2}+8\right) \\
&  &
+ \tau ^3 \frac{16 \left(A^2+1\right) (k+1)}{\left(A^2-1\right)^2} \left(\frac{(k+2)^2 \Gamma (-2 k-3)}{\Gamma (-k)^2}+1\right)+O(\tau ^4)\,.
\end{eqnarray*}
Hence
\begin{eqnarray*} 
&  &  
\left(1+x \right) \left(\left(1+\tau  \right)^2-A^2 \right)  
F \left(k+1,k+1;1;z\right)
+(2k+2) \left(1-\tau ^{ 2 }-A^2 \right)  
F \left(k+2,k+2;2;z\right)\\
&  = & 
2 (k+1) \tau  \left(\frac{\left(A^2 (-(2 k+3))+12 k+19\right) \Gamma (-2 k-3)}{\Gamma (-k)^2}+4\right)\\
&  &
+\tau ^2 \frac{4 }{A^2-1} \left(4 k-\frac{(k+1) \left(A^2 \left(2 k^2+k-3\right)-k (6 k+17)-13\right) \Gamma (-2 k-3)}{\Gamma (-k)^2}\right)\\
&  &
+\tau ^3 \frac{2}{\left(A^2-1\right)^2}   \left(8 \left(A^2+1\right) k+\frac{(k+1) \left(A^2+4 k+7\right) \left(A^2 (2 k+3)+2 k+5\right) \Gamma (-2 k-3)}{\Gamma (-k)^2}\right)
+O(\tau )^4\,.
\end{eqnarray*}
Here the term with $ \tau $ is 
\begin{eqnarray*} 
&  &  
2 \frac{(k+1)}{\Gamma (-k)^2}  \left[ (A^2 (- 2 k-3 )+12 k+19 ) \Gamma (-2 k-3)+4\Gamma (-k)^2\right]\tau  
\end{eqnarray*}
and for $k=-2$ it takes value $2 \frac{(k+1)}{\Gamma (-k)^2}   (A^2-1)\tau $.
Corollary is proved. \qed
\medskip

\noindent
Now we can complete the proof of   necessity part in Theorem~\ref{T0.3} for the case of $\ell=2k+1 $, $k=   -3, -4,\ldots $.  Indeed, for these values of $\ell $ for the component $\Psi_0 $ of the solution we have 
\begin{eqnarray*} 
\Psi_0 (0,t) 
& = &
   2 e^{-Ht}  
  \int_{0}^{1 } \left( \frac{\partial }{\partial s} s\Phi _0 ( s) \right) \Bigg\{ 
H  2^\ell (1+\ell )  e^{-\frac{1}{2} H (\ell+2) t} \left(\left(1+e^{-H t}\right)^2-H^2 r^2\right)^{-\frac{1}{2} (\ell+5)} \\
&  &
\times \Bigg[\left(1+e^{-H t} \right) \left(\left(1+e^{-H t} \right)^2-H^2 r^2\right)  
F \left(\frac{\ell+1}{2},\frac{\ell+1}{2};1;\frac{\left(1-e^{-H t}\right)^2-H^2 r^2}{\left(1+e^{-H t}\right)^2-H^2 r^2}\right)\\
&  &
+(\ell+1) \left(1-e^{-2 H t}-H^2 r^2 \right)  
F \left(\frac{\ell+3}{2},\frac{\ell+3}{2};2;\frac{\left(1-e^{-H t}\right)^2-H^2 r^2}{\left(1+e^{-H t}\right)^2-H^2 r^2}\right)\Bigg]\Bigg\}   ds  \\
& = &
  e^{-2Ht}  H  2^{2k+2} (2+ 2k  )^2  e^{-\frac{1}{2} H (2k+3) t}
  \int_{0}^{1 } \left( \frac{\partial }{\partial r}r\Phi _0 ( r) \right) \Bigg\{ 
 \left(\left(1+e^{-H t}\right)^2-H^2 r^2\right)^{-  (k+3)} \\
&  &
\times  \Bigg[ \left(\frac{\left(H^2 r^2 (- 2 k-3 )+12 k+19\right) \Gamma (-2 k-3)+4\Gamma (-k)^2}{\Gamma (-k)^2}\right)
+e^{- H t} O(1)\Bigg]\Bigg\}   dr  \,.
\end{eqnarray*}
Next for every given $k=-3,-4,\ldots$, ($\ell=2k+1 $) we choose the function $ \Phi _0$ such that 
\begin{equation} 
\label{6.23}
  \int_{0}^{1 } \left( \frac{\partial }{\partial r}r\Phi _0 ( r) \right) 
 \left(1-H^2 r^2\right)^{-  (k+3)}  \left[ (H^2 r^2  (- 2 k-3 )+12 k+19 ) \Gamma (-2 k-3)+4\Gamma (-k)^2\right]  dr\not=0  
\end{equation} 
for all values of $H  r =A $ from Corollary~\ref{C5.2}. Here we note that only for $k=-2$ this integral is reduced to the expression that proportional to  
\begin{eqnarray*} 
&  &
  \int_{0}^{1 } \left( \frac{\partial }{\partial r}r\Phi _0 ( r) \right) dr= 0 . 
\end{eqnarray*} 
For all values $k=-3,-4,\ldots$, 
the   equation (\ref{6.23}) shows that the value  $\Psi_1 (0,t)$ for large time depends on the values of initial function inside of the characteristic conoid. This completes the proof in this case.

\subsection{The case of positive odd $\ell=2k+1 $, $ k =1,2,\ldots$}

We skip the proof of the following simple lemma.
\begin{lemma} 
\label{L5.6}
There are the following limits
\begin{eqnarray*} 
  \lim_{x\searrow  0} x^{2 n-1} F (n,n;1;1-x)
& = &
 F (1-n,1-n;1;1)=\frac{\Gamma (n-1)}{[\Gamma (n)]^2}, \quad n=2,3,\ldots\,,\\
  \lim_{x\searrow  0} x^{2 n-2} F (n,n;2;1-x)
& = &
 F (2-n,2-n;2;1)=\frac{\Gamma (2n-2)}{[\Gamma (n)]^2}, \quad n=2,3,\ldots\,.\\
\end{eqnarray*}
\end{lemma}

\begin{corollary}
\label{C5.4}
For $\ell=2k+1 $, $ k =1,2,\ldots$  
\begin{eqnarray*} 
& &
{\mathcal F} \left ( e^{-H t},H  r ;\ell;\frac{\left(1-e^{-H t}\right)^2-H^2 r^2}{\left(1+e^{-H t}\right)^2-H^2 r^2} \right)\\
& = & 
(2k+2) \left(1 -H^2 r^2 \right)^{3+2k}    4  ^{ -2 k-2 }    e^{ H(2k+2) t} 
\Bigg\{\frac{ \Gamma (2 k+2 )}{[\Gamma (k+2)]^2}  
+R  (k,H,r;t) \Bigg\}
\,,\quad k =1,2,\ldots\,,
\end{eqnarray*}
where  with  large  $T$  the remainder  $R (k,H,r;t) $ can be estimated as follows 
\[
|R (k,H,r;t)| \leq o(1)  \quad \mbox{\rm for all} \quad  t \geq T \quad  and  \quad   0 \leq r \leq 1/(2H) \,.
\]
\end{corollary}
\medskip

\noindent
{\bf Proof.}
From Lemma~\ref{L5.6}, with $n=k+1>1$ and $n=k+2$ ($ k >0$) we obtain 
\begin{eqnarray*} 
&  &  
\left(1+\tau  \right) \left(\left(1+\tau \right)^2-H^2 r^2\right)  
F \left(k+1,k+1;1;z\right)
+(2k+2) \left(1-\tau ^2-A^2 \right)  
F \left(k+2,k+2;2;z\right)\\
&  = & 
\left(1-A^2\right) \left(1-z\right)^{1-2n}   
\left[ \frac{\Gamma (n-1)}{[\Gamma (n)]^2}+R_1 (k,H,r;t) \right] \\
&  &
+(2k+2) \left(1 -A^2 \right)  \left(1-z\right)^{2-2n}
\left[ \frac{\Gamma (2n-2)}{[\Gamma (n)]^2}  +R_2 (k,H,r;t)\right]\\ 
&  = & 
\left(1-A^2\right) \left(1-z\right)^{ -2 k-1 }   
 \left[ \frac{\Gamma (k)}{[\Gamma  (k+1) ]^2}+R_1 (k,H,r;t) \right]\\
&  &
+(2k+2) \left(1 -A^2 \right)  \left(1-z\right)^{ -2 k-2 }
\left[ \frac{\Gamma (2 k+2 )}{[\Gamma (k+2)]^2}  
+R_2 (k,H,r;t) \right]\,.
\end{eqnarray*}
It follows
\begin{eqnarray*} 
&  &  
\left(1+\tau  \right) \left(\left(1+\tau \right)^2-H^2 r^2\right)  
F \left(k+1,k+1;1;z\right)
+(2k+2) \left(1-\tau ^2-A^2 \right)  
F \left(k+2,k+2;2;z\right)\\
&  = & 
\left(1-A^2\right)^{2+2k}    4  ^{ -2 k-1 }    e^{ H(2k+1) t}   
 \left[ \frac{\Gamma (k)}{[\Gamma  (k+1) ]^2}+R_1 (k,H,r;t) \right]\\
&  &
+(2k+2) \left(1 -A^2 \right)^{3+2k}    4  ^{ -2 k-2 }    e^{ H(2k+2) t} 
\left[ \frac{\Gamma (2 k+2 )}{[\Gamma (k+2)]^2}  
+R_2 (k,H,r;t)  \right]\\
&  = & (2k+2) \left(1 -H^2 r^2 \right)^{3+2k}    4  ^{ -2 k-2 }    e^{ H(2k+2) t} 
\left[ \frac{ \Gamma (2 k+2 )}{[\Gamma (k+2)]^2}  
+R  (k,H,r;t)  \right]
\,,\quad k=1,2,\ldots\,.
\end{eqnarray*}
Finally, since $ k >0$  we obtain the statement of corollary. \hfill 
$\square$
\medskip

\noindent
 In order to complete the proof of (i) Theorem~\ref{T0.3} for the case of $\ell=2k+1 $, $k=1,2 ,3, \ldots $  we apply  Corollary~\ref{C5.4}   and write
\begin{eqnarray*} 
\Psi_1 (0,t) 
& = &
   2 e^{-Ht}  H  2^\ell (1+\ell )  e^{-\frac{1}{2} H (\ell+2) t}
  \int_{0}^{1 } \left( \frac{\partial }{\partial s} s\Phi _0 ( s) \right) 
 \left(\left(1+e^{-H t}\right)^2-H^2 r^2\right)^{-\frac{1}{2} (\ell+5)} \\
&  &
\times (2k+2) \left(1 -H^2 r^2 \right)^{3+2k}    4  ^{ -2 k-2 }    e^{ H(2k+2) t} 
\left[ \frac{ \Gamma (2 k+2 )}{[\Gamma (k+2)]^2}  
+R  (k,H,r;t) \right]   ds  \\
& = & 
   2^{-2k-2}   H     (2k+2)^2 e^{ \frac{1}{2} H (2k-1) t}
  \int_{0}^{1 } \left( \frac{\partial }{\partial r} r\Phi _0 ( r) \right) 
 \left(\left(1+e^{-H t}\right)^2-H^2 r^2\right)^{-(k+3)} \\
&  &
\times  \left(1 -H^2 r^2 \right)^{3+2k}         
\left[ \frac{ \Gamma (2 k+2 )}{[\Gamma (k+2)]^2}  
+R  (k,H,r;t) \right]  \,   dr   \,.
\end{eqnarray*}
For every given $ k=1,2 ,3, \ldots $ we  choose $ \Phi _0$ such that
\begin{eqnarray*}  
&   & 
  \int_{0}^{1 } \left( \frac{\partial }{\partial r} r\Phi _0 ( r) \right) 
 \left(1-H^2 r^2\right)^{\frac{1}{2}(\ell-1)}    dr =  \int_{0}^{1 } \left( \frac{\partial }{\partial r} r\Phi _0 ( r) \right) 
 \left(1-H^2 r^2\right)^{k}    dr   \not= 0 .
\end{eqnarray*}
The last equation shows that the value  $\Psi_1 (0,t)$ for large time depends on the values of initial function inside of the characteristic conoid. This completes the proof in this case.

\section{Necessity of $m=0,\pm iH$  for the Huygens  principle. The case of  even $\ell=2k $, $k=0,\pm 1, \pm2,\ldots $}
\label{S6}

For the case of $\ell=2k $, $k=0,\pm 1, \pm2,\ldots $ we have
\begin{eqnarray*}
{\mathcal F} \left ( e^{-H t},H  r ;\ell;\frac{\left(1-e^{-H t}\right)^2-H^2 r^2}{\left(1+e^{-H t}\right)^2-H^2 r^2} \right) 
& = &  
\left(1+\tau  \right) \left(\left(1+\tau  \right)^2-A^2\right)  
F \left(k+\frac{1}{2},k+\frac{1}{2};1;z\right)\\
&  &
+(2k+1) \left(1-\tau ^{2}-A^2 \right)  
F \left(k+\frac{3}{2},k+\frac{3}{2};2;z\right)  \,.
\end{eqnarray*} 

\subsection{The case of positive even $\ell=2k $, $k=1,2,3,\ldots$}

If $k=1,2,3,\ldots$ and $c-a-b=-2k$,  then we apply  (\ref{mBEA}) (\cite[(14) Sec. 2.10]{B-E}). 
Hence,
\begin{eqnarray*}  
\lim_{z \nearrow  1} F \left(k+\frac{1}{2},k+\frac{1}{2};1;z\right) 
& = &
\frac{\Gamma(2k)(1-z)^{-2k}}{ [\Gamma(k+\frac{1}{2})]^2} \sum_{n=0}^{2k-1} \frac{[( \frac{1}{2}- k)_{n}]^2}{(1-2k)_{n} n !}(1-z)^{n}   \nonumber \\
&  &
+\frac{(-1)^ {2k} }{[\Gamma(k+\frac{1}{2}-2k)]^2} \sum_{n=0}^{\infty} \frac{[(k+\frac{1}{2})_{n}]^2}{(n+2k)_{n} n !}\left[\bar{h}_{n}-\ln (1-z)\right](1-z)^{n}\,,
\end{eqnarray*} 
where $\bar{h}_{n}=\psi(1+n)+\psi(1+n+2k)-\psi(a+n)-\psi(b+n)$, 
implies 
\begin{eqnarray*} 
F \left(k+\frac{1}{2},k+\frac{1}{2};1;z\right)
& = &
\frac{\Gamma(2k)(1-z)^{-2k}}{ [\Gamma(k+\frac{1}{2})]^2}     
+\frac{\Gamma(2k)(1-z)^{-2k}}{ [\Gamma(k+\frac{1}{2})]^2} \sum_{n=1}^{2k-1} \frac{[( \frac{1}{2}- k)_{n}]^2}{(1-2k)_{n} n !}(1-z)^{n} \nonumber \\
&  &
+\frac{(-1)^ {2k} }{[\Gamma(k+\frac{1}{2}-2k)]^2} \sum_{n=0}^{\infty} \frac{[(k+\frac{1}{2})_{n}]^2}{(n+2k)_{n} n !}\left[\bar{h}_{n}-\ln (1-z)\right](1-z)^{n} \\
& = &
\frac{\Gamma(2k)(1-z)^{-2k}}{ [\Gamma(k+\frac{1}{2})]^2}      
+(1-z)^{-2k+1} O(1) 
+|\ln (1-z)|  O(1),\quad  k=1,2,\ldots\,,
\end{eqnarray*} 
and
\begin{eqnarray*} 
 (1-z)^{2k} F \left(k+\frac{1}{2},k+\frac{1}{2};1;z\right)
& = &
\frac{\Gamma(2k)}{ [\Gamma(k+\frac{1}{2})]^2}  
+(1-z)  O(1) ,\quad  k=1,2,\ldots\,,\\
\lim_{z \nearrow  1} (1-z)^{2k} F \left(k+\frac{1}{2},k+\frac{1}{2};1;z\right)
& = &
\frac{\Gamma(2k)}{ [\Gamma(k+\frac{1}{2})]^2}   ,\quad  k=1,2,\ldots\,. 
\end{eqnarray*} 
Further, according to  (\ref{ssA3})  with $m= 2k+1$, for $k=1,2,\ldots$ we have
\begin{eqnarray*} 
&  &
 F \left(k+\frac{3}{2},k+\frac{3}{2};2;z\right)\\
& = &
\frac{\Gamma(2k+1)(1-z)^{-(2k+1)}}{[\Gamma(k+\frac{3}{2}) ]^2} \sum_{n=0}^{(2k+1)-1} 
\frac{[( \frac{1}{2}- k )_{n}]^2}{( - 2k )_{n} n !}(1-z)^{n}   \nonumber \\
&  &
+\frac{(-1)^{(2k+1)}}{[\Gamma(\frac{1}{2}- k)]^2} \sum_{n=0}^{\infty} \frac{[(k+\frac{3}{2})_{n}]^2}{(n+ 2k+1 )_{n} n !}\left[\bar{h}_{n}-\ln (1-z)\right](1-z)^{n} \\ 
& = &
\frac{\Gamma(2k+1)(1-z)^{-(2k+1)}}{[\Gamma(k+\frac{3}{2}) ]^2}  
+\frac{\Gamma(2k+1)(1-z)^{-(2k+1)}}{[\Gamma(k+\frac{3}{2}) ]^2} \sum_{n=1}^{(2k+1)-1} 
\frac{[( \frac{1}{2}- k )_{n}]^2}{( - 2k )_{n} n !}(1-z)^{n}   \nonumber \\
&  &
+\frac{(-1)^{(2k+1)}}{[\Gamma( \frac{1}{2}- k )]^2} \sum_{n=0}^{\infty} \frac{[(k+\frac{3}{2})_{n}]^2}{(n+ 2k+1 )_{n} n !}\left[\bar{h}_{n}-\ln (1-z)\right](1-z)^{n} \\ 
& = &
\frac{\Gamma(2k+1)(1-z)^{-(2k+1)}}{[\Gamma(k+\frac{3}{2}) ]^2}  
+\frac{\Gamma(2k+1)}{[\Gamma(k+\frac{3}{2}) ]^2}(1-z)^{- 2k } O(1)  
+|\ln (1-z)|O(1)    \,,
\end{eqnarray*} 
which implies
\begin{eqnarray*}
(1-z)^{ (2k+1)} F \left(k+\frac{3}{2},k+\frac{3}{2};2;z\right)
& = &
\frac{\Gamma(2k+1)}{[\Gamma(k+\frac{3}{2}) ]^2}
+ (1-z)  O(1) \,,  \\  
\lim_{z \nearrow  1}(1-z)^{ (2k+1)} F \left(k+\frac{3}{2},k+\frac{3}{2};2;z\right)
& = &
\frac{\Gamma(2k+1)}{[\Gamma(k+\frac{3}{2}) ]^2}\,.
\end{eqnarray*} 
Hence,
\begin{eqnarray*} 
&  &
\left(1+\tau  \right) \left(\left(1+\tau  \right)^2-A^2\right)  
F \left(k+\frac{1}{2},k+\frac{1}{2};1;\frac{\left(1-x\right)^2-A^2}{\left(1+x\right)^2-A^2}\right)\\
&  &
+(2k+1) \left(1-\tau ^{2}-A^2 \right)  
F \left(k+\frac{3}{2},k+\frac{3}{2};2;\frac{\left(1-x\right)^2-A^2}{\left(1+x\right)^2-A^2}\right) \\
& = &
\left(1+\tau  \right) \left(\left(1+\tau  \right)^2-A^2\right)
 (1-z)^{-2k} \left\{ \frac{\Gamma(2k)}{ [\Gamma(k+\frac{1}{2})]^2}  
+(1-z)  O(1) \right\}\\
&  &
+(2k+1) \left(1-\tau ^{2}-A^2 \right)(1-z)^{- (2k+1)}\left\{ \frac{\Gamma(2k+1)}{[\Gamma(k+\frac{3}{2}) ]^2}
+ (1-z)  O(1)   \right\}  \\
& = &
(2k+1) \left(1-\tau ^{2}-A^2 \right)(1-z)^{- (2k+1)}\left\{ \frac{\Gamma(2k+1)}{[\Gamma(k+\frac{3}{2}) ]^2}
+ (1-z)  O(1)   \right\}  \,.
\end{eqnarray*} 
In particular,
\begin{eqnarray*} 
&  &
\lim_{\tau  \to 0} (1-z)^{  (2k+1)}\Bigg[\left(1+\tau  \right) \left(\left(1+\tau  \right)^2-A^2\right)  
F \left(k+\frac{1}{2},k+\frac{1}{2};1;z\right)\\
&  &
+(2k+1) \left(1-\tau ^{2}-A^2 \right)  
F \left(k+\frac{3}{2},k+\frac{3}{2};2;z\right) \Bigg]  \nonumber \\
& = &
(2k+1) 
\frac{\Gamma(2k+1)}{[\Gamma(k+\frac{3}{2}) ]^2} \left(1-A^2 \right), \quad k=1,2,3,\ldots \nonumber  \,.
\end{eqnarray*} 
\medskip

\noindent
{In order to complete the  proof of  necessity part in Theorem~\ref{T0.3} for $\ell=2k $, $k=1,2,3,\ldots$} 
we write
\begin{eqnarray*} 
\Psi_1 (0,t) 
& = &
   2 e^{-Ht}  
  \int_{0}^{1 } \left( \frac{\partial }{\partial r} r\Phi _0 (r) \right) \Bigg\{ 
H  2^\ell (1+\ell )  e^{-\frac{1}{2} H (\ell+2) t} \left(\left(1+e^{-H t}\right)^2-H^2 r^2\right)^{-\frac{1}{2} (\ell+5)} \\
&  &
\times \Bigg\{\left(1+e^{-H t} \right) \left(\left(1+e^{-H t} \right)^2-H^2 r^2\right)  
F \left(\frac{\ell+1}{2},\frac{\ell+1}{2};1;\frac{\left(1-e^{-H t}\right)^2-H^2 r^2}{\left(1+e^{-H t}\right)^2-H^2 r^2}\right)\\
&  &
+(\ell+1) \left(1-e^{-2 H t}-H^2 r^2 \right)  
F \left(\frac{\ell+3}{2},\frac{\ell+3}{2};2;\frac{\left(1-e^{-H t}\right)^2-H^2 r^2}{\left(1+e^{-H t}\right)^2-H^2 r^2}\right) \Bigg\} \Bigg\} dr  \\ 
& = &
   2 e^{-Ht}  
  \int_{0}^{1 } \left( \frac{\partial }{\partial r} r\Phi _0 (r) \right)  
H  2^\ell (1+\ell )  e^{-\frac{1}{2} H (\ell+2) t} \left(\left(1+e^{-H t}\right)^2-H^2 r^2\right)^{-\frac{1}{2} (\ell+5)} \\
&  &
\times (2k+1) \left(1-\tau ^{2}-A^2 \right)\left( \frac{4e^{-H t} }{\left(1+e^{-H t}\right)^2-H^2 r^2}\right)^{- (2k+1)}\left[ \frac{\Gamma(2k+1)}{[\Gamma(k+\frac{3}{2}) ]^2}
+   o(1)   \right]  dr  \\ 
& = &
    H  2^{-2k-1}e^{Ht(k-1)}  (2k+1)^2
  \int_{0}^{1 } \left( \frac{\partial }{\partial r} r\Phi _0 (r) \right)  
     \left(\left(1+e^{-H t}\right)^2-H^2 r^2\right)^{ \frac{1}{2} (2k-3)} \\
&  &
\times  \left(1-e^{-2Ht}-H^2 r^2 \right)  \left[ \frac{\Gamma(2k+1)}{[\Gamma(k+\frac{3}{2}) ]^2}
+   o(1)   \right]   dr\,.
\end{eqnarray*} 
Next for every $\ell=2k $,   $k=1,2,3,\ldots $ we can we choose a function $\Phi _0 $ such that
\begin{eqnarray*}
&   &
  \int_{0}^{1 } \left( \frac{\partial }{\partial r} r\Phi _0 (r) \right)  
     \left(1-H^2 r^2\right)^{ \frac{1}{2} (\ell-1)}   dr =  \int_{0}^{1 } \left( \frac{\partial }{\partial r} r\Phi _0 (r) \right)  
     \left(1-H^2 r^2\right)^{ \frac{1}{2} (2k-1)}   dr\not= 0  \,.
\end{eqnarray*}
The last equation shows that the value  $\Psi_1 (0,t)$ for large time depends on the values of initial function inside of the characteristic conoid. This completes the proof in this case.

\subsection{The case of negative even  $\ell=2k $, $k=-2,-3,\ldots$}

For $\ell=2k $, $k=-2,-3,\ldots$ consider the function
\begin{eqnarray*}
{\mathcal F} \left ( e^{-H t},H  r ;\ell;\frac{\left(1-e^{-H t}\right)^2-H^2 r^2}{\left(1+e^{-H t}\right)^2-H^2 r^2} \right) 
& = &   
\left(1+\tau  \right) \left(\left(1+\tau  \right)^2-A^2\right)  
F \left(k+\frac{1}{2},k+\frac{1}{2};1;\frac{\left(1-\tau \right)^2-A^2}{\left(1+\tau \right)^2-A^2}\right)\\
&  &
+(2k+1) \left(1-\tau ^{2}-A^2 \right)  
F \left(k+\frac{3}{2},k+\frac{3}{2};2;\frac{\left(1-\tau \right)^2-A^2}{\left(1+\tau \right)^2-A^2}\right) \,. 
\end{eqnarray*} 
\begin{lemma}
\label{L7.1}
The following formulas with $z\searrow 0 $ hold
\begin{eqnarray*}  
F \left(k+\frac{1}{2},k+\frac{1}{2};1;1-z\right) 
& = &
\frac{ \Gamma (-2k)}{[\Gamma (\frac{1}{2}-k)]^2}-z \left(k+\frac{1}{2}\right)^2  \frac{ \Gamma (-1-2k)}{[\Gamma (\frac{1}{2}-k)]^2}+O(z^2),\\
&  &
\hspace{4cm}  k= -1, -2,-3,\ldots\,,\\ 
F \left(k+\frac{3}{2},k+\frac{3}{2};2;1-z\right)  
& = &
\frac{ \Gamma (-1-2k)}{[\Gamma (\frac{1}{2}-k)]^2}
-z    \left(k+\frac{3}{2}\right)^2   \frac{  \Gamma (-2-2k)}{[\Gamma (\frac{1}{2}-k)]^2} +O(z^2) ,\\
&  &
\hspace{4cm}  k=  -2,-3,\ldots\,.
\end{eqnarray*}
\end{lemma}
\medskip

\noindent
{\bf Proof.}
For the function $F \left(k+\frac{1}{2},k+\frac{1}{2};1;z\right) $, since $c-a-b=-2k>0 $ and   $c- b=\frac{1}{2}-k >0 $,  we apply \cite[(4) Sec. 2.1]{B-E}
\begin{eqnarray*}  
F \left(k+\frac{1}{2},k+\frac{1}{2};1;1\right) 
& = & 
\frac{\Gamma (1)\Gamma (-2k)}{[\Gamma (\frac{1}{2}-k)]^2},\quad k=-1,-2,\ldots\,.
\end{eqnarray*} 
For the function $F \left(k+\frac{3}{2},k+\frac{3}{2};2;z\right)$, since $c-a-b=-1-2k>0 $ and   
$c- b=\frac{1}{2}-k >0 $  we apply \cite[(4) Sec. 2.1]{B-E}
\begin{eqnarray}  
\label{8.29}
F \left(k+\frac{3}{2},k+\frac{3}{2};2;1\right)
& = & 
\frac{\Gamma (2)\Gamma (-1-2k)}{[\Gamma (\frac{1}{2}-k)]^2},\quad k=-1,-2,\ldots\,.
\end{eqnarray} 
It follows
\begin{eqnarray*} 
F \left(k+\frac{1}{2},k+\frac{1}{2};1;1\right) 
+(2k+1)  
F \left(k+\frac{3}{2},k+\frac{3}{2};2;1\right)  
& = &
\frac{ \Gamma (-2k)}{[\Gamma (\frac{1}{2}-k)]^2}
+(2k+1)  
\frac{ \Gamma (-1-2k)}{[\Gamma (\frac{1}{2}-k)]^2} \\
& = &
0,\quad k=-1,-2,\ldots\,.
\end{eqnarray*} 
Then we look at the next term
\begin{eqnarray*}  
\left[ \frac{d}{dz}F \left(k+\frac{1}{2},k+\frac{1}{2};1;z\right) \right]_{z=1}
& = & 
-\left[ \frac{d}{dz}F \left(k+\frac{1}{2},k+\frac{1}{2};1;1-z\right) \right]_{z=0} \\
& = & 
-\left[ -\left(k+\frac{1}{2}\right)^2  F \left(k+\frac{3}{2},k+\frac{3}{2};2;1-z\right) \right]_{z=0} \\
& = & 
  \left(k+\frac{1}{2}\right)^2  \frac{ \Gamma (-1-2k)}{[\Gamma (\frac{1}{2}-k)]^2}  ,\quad k=-1,-2,\ldots\,.
\end{eqnarray*} 
Also
\begin{eqnarray*}  
\left[ \frac{d}{dz}F \left(k+\frac{3}{2},k+\frac{3}{2};2;z\right) \right]_{z=1}
& = & 
-\left[ \frac{d}{dz}F \left(k+\frac{3}{2},k+\frac{3}{2};2;1-z\right) \right]_{z=0} \\
& = & 
 \frac{1}{2} \left(k+\frac{3}{2}\right)^2  F \left(k+\frac{5}{2},k+\frac{5}{2};3;1 \right)\\
& = & 
 \frac{1}{2} \left(k+\frac{3}{2}\right)^2   \frac{ 2\Gamma (-2-2k)}{[\Gamma (\frac{1}{2}-k)]^2}  ,\quad k=  -2,-3,\ldots\,.
\end{eqnarray*}
It follows
\begin{eqnarray*} 
&  & 
F \left(k+\frac{1}{2},k+\frac{1}{2};1;1\right) 
+(2k+1)  
F \left(k+\frac{3}{2},k+\frac{3}{2};2;1\right) \\
& = &
 \left(k+\frac{1}{2}\right)^2  \frac{ \Gamma (-1-2k)}{[\Gamma (\frac{1}{2}-k)]^2}
+(2k+1)  
\frac{1}{2} \left(k+\frac{3}{2}\right)^2   \frac{ 2\Gamma (-2-2k)}{[\Gamma (\frac{1}{2}-k)]^2} \\
& = &
\frac{(6 k+7) \Gamma (-2 k)}{8 (k+1)[\Gamma (\frac{1}{2}-k)]^2} >0,\quad k= -2,-3,\ldots\,.
\end{eqnarray*}
The lemma is proved. \hfill $\square$

 The case of $k=-1 $, that is,  $\ell =-2$ and $ m =  -  i\frac{H}{2}  $ will be discussed in subsection~\ref{SS7.3}. 
\begin{corollary}
For all $k=-2,-3,\ldots$ and $0\leq A<1/2$  
\begin{eqnarray*} 
&  &
\left(1+\tau  \right) \left(\left(1+\tau  \right)^2-A^2\right)  
F \left(k+\frac{1}{2},k+\frac{1}{2};1;\frac{\left(1-\tau \right)^2-A^2}{\left(1+\tau \right)^2-A^2}\right)\\
&  &
+(2k+1) \left(1-\tau ^{2}-A^2 \right)  
F \left(k+\frac{3}{2},k+\frac{3}{2};2;\frac{\left(1-\tau \right)^2-A^2}{\left(1+\tau \right)^2-A^2}\right) \\
& = &
\frac{\left(-2 A^2 (k+1)+2 k+1\right) \Gamma (-2 k)}{2 (k+1) \Gamma \left(\frac{1}{2}-k\right)^2} \tau +O\left(\tau ^2\right) \,,
\end{eqnarray*}
where
\[
\frac{\left(-2 A^2 (k+1)+2 k+1\right) \Gamma (-2 k)}{2 (k+1) \Gamma \left(\frac{1}{2}-k\right)^2}>0, \quad k=-2,-3,\ldots\,.
\]
\end{corollary}
\medskip

\noindent
{\bf Proof.} Indeed, according to Lemma~\ref{L7.1}
\begin{eqnarray*} 
&  &
\left(1+\tau  \right) \left(\left(1+\tau  \right)^2-A^2\right)  
F \left(k+\frac{1}{2},k+\frac{1}{2};1;\frac{\left(1-\tau \right)^2-A^2}{\left(1+\tau \right)^2-A^2}\right)\\
&  &
+(2k+1) \left(1-\tau ^{2}-A^2 \right)  
F \left(k+\frac{3}{2},k+\frac{3}{2};2;\frac{\left(1-\tau \right)^2-A^2}{\left(1+\tau \right)^2-A^2}\right) \\
& = &
(1-A^2  +(3-A^2) \tau +O(\tau ^2))\\
&  &
\times \Bigg[ \frac{ \Gamma (-2k)}{[\Gamma (\frac{1}{2}-k)]^2}+\left(\frac{4 \tau }{1-A^2}+\frac{8 \tau ^2}{\left(A^2-1\right)^2}+O\left(\tau ^3\right) \right) \left(k+\frac{1}{2}\right)^2  \frac{ \Gamma (-1-2k)}{[\Gamma (\frac{1}{2}-k)]^2}+O(\tau ^2)\Bigg]\\
&  &
+(2k+1) \left(1-A^2-x^{2} \right) \\
&  &
\times \Bigg[ \frac{ \Gamma (-1-2k)}{[\Gamma (\frac{1}{2}-k)]^2}
-\left(\frac{4 \tau }{1-A^2}+\frac{8 \tau ^2}{\left(A^2-1\right)^2}+O\left(x^3\right)\right)    \left(k+\frac{3}{2}\right)^2   \frac{  \Gamma (-2-2k)}{[\Gamma (\frac{1}{2}-k)]^2} +O(z^2) \Bigg]\\
& = &
\frac{\left(-2 A^2 (k+1)+2 k+1\right) \Gamma (-2 k)}{2 (k+1) \Gamma \left(\frac{1}{2}-k\right)^2} \tau +O\left(\tau ^2\right) 
\end{eqnarray*}
provided that $k=-2,-3,\ldots$ and 
$
0\leq   A^2 < 1/4  
$. 
The corollary is proved. \hfill $\square$
\medskip

\noindent
 To complete the proof of (i) Theorem~\ref{T0.3} for the case of $\ell=2k $, $k=-2,-3,\ldots$  
we write
\begin{eqnarray*}
\Psi_0 (0,t) 
& = &
     H  2^{2k+1} (1+2k )  
e^{-  H ( k+3) t}  
  \int_{0}^{1 } \left( \frac{\partial }{\partial s} s\Phi _0 ( s) \right)   \left(\left(1+e^{-H t}\right)^2-H^2 r^2\right)^{-\frac{1}{2} (2k+5)} \\
&  &
\times \Bigg[\frac{\left(-2 H^2 r^2 (k+1)+2 k+1\right) \Gamma (-2 k)}{2 (k+1) \Gamma \left(\frac{1}{2}-k\right)^2}  +e^{-  H t}O\left(1\right) \Bigg]   ds  \,.
\end{eqnarray*}
Next for every $k=-2,-3,\ldots$ we can chose choose $\Phi _0 $ such that 
\begin{eqnarray*} 
& & 
  \int_{0}^{1 } \left( \frac{\partial }{\partial r} r\Phi _0 ( r) \right)   \left(1-H^2 r^2\right)^{-\frac{1}{2} (2k+5)}\left(-2 H^2 r^2 (k+1)+2 k+1\right)  dr  \not= 0\,.
\end{eqnarray*}
The last equation shows that the value  $\Psi_0 (0,t)$ for large time depends on the values of initial function inside of the characteristic conoid. This completes the proof in this case. 
\hfill $\square$

\subsection{The case of $ m =  -  i\frac{H}{2}  $,\, that is
$\ell=2k $, $k=-1 $, $\ell =-2 $ }
\label{SS7.3}

{For $\ell=2k $, $k=-1 $,}  consider
\begin{eqnarray*}  
F \left(k+\frac{3}{2},k+\frac{3}{2};2;1-z\right)  
& = & 
F \left( \frac{1}{2},\frac{1}{2};2;1-z\right)\,.  
\end{eqnarray*}
We have $c-a-b=1$ and $c-a= \frac{3}{2}  $ and apply (\ref{8.29}):
\begin{eqnarray*}  
F \left(k+\frac{3}{2},k+\frac{3}{2};2;1 \right)  
& = & 
F \left( \frac{1}{2},\frac{1}{2};2;1 \right)  = \frac{ 1}{[\Gamma (\frac{3}{2} )]^2}=\frac{4}{ \pi}\,,
\end{eqnarray*}
while
\begin{eqnarray*}  
  \frac{d}{dz}F \left(k+\frac{3}{2},k+\frac{3}{2};2;z\right)  
& = & 
\frac{1}{8}  F \left(\frac{3}{2},\frac{3}{2};3; z\right) \,.
\end{eqnarray*}
Then we use (\ref{A2}) with $m=0$, that is,
\begin{eqnarray*} 
F \left(\frac{3}{2},\frac{3}{2};3; z\right) 
& = &
\frac{\Gamma (3)}{[\Gamma (\frac{3}{2})]^2}
\left[ 2\psi ( 1) - 2\psi \left(\frac{3}{2} \right) - \ln (1-z) \right]  \\
&  &
+\frac{\Gamma (3)}{[\Gamma (\frac{3}{2})]^2}\sum _{n=1}^\infty \frac{[(\frac{3}{2})_n]^2}{(n!)^2}
\left[ 2\psi (n+1) - 2\psi  \left(\frac{3}{2}+n\right) - \ln (1-z) \right] (1-z)^{n}\\
& = &
\frac{8}{\pi} 
\left[ 2\psi ( 1) - 2\psi \left(\frac{3}{2} \right) - \ln (1-z) \right]   
+O(1) (1-z) \ln (1-z)\,,
\nonumber  \\
 &  &
\hspace*{5cm} \quad |\arg (1-z) |<\pi , \quad |1-z|<1 \,.\nonumber  
\end{eqnarray*}
Hence
\begin{eqnarray*}  
 \frac{d}{dz}F \left(k+\frac{3}{2},k+\frac{3}{2};2;z\right)  
& = & 
  \frac{1}{  {\pi } }  
\left[ 2\psi ( 1) - 2\psi \left(\frac{3}{2} \right) - \ln (1-z) \right]   
+O(1) (1-z) \ln (1-z) \,.
\end{eqnarray*}
It follows for $k=-1$
\begin{eqnarray*} 
&  &
(\tau +1) \left((\tau +1)^2-A^2\right)  F \left(k+\frac{1}{2},k+\frac{1}{2};1;\frac{(1-\tau )^2-A^2}{(\tau +1)^2-A^2}\right)\\
&  &
+(2 k+1) \left(-A^2-\tau +1\right)  F \left(k+\frac{3}{2},k+\frac{3}{2};2;\frac{(1-\tau )^2-A^2}{(\tau +1)^2-A^2}\right)\\
& = &
(\tau +1) \left((\tau +1)^2-A^2\right)  F \left(-\frac{1}{2},-\frac{1}{2};1;\frac{(1-\tau )^2-A^2}{(\tau +1)^2-A^2}\right)\\
&  &
- \left(-A^2-\tau +1\right)  F \left( \frac{1}{2}, \frac{1}{2};2;\frac{(1-\tau )^2-A^2}{(\tau +1)^2-A^2}\right)\,.
\end{eqnarray*}
According to Lemma~\ref{L7.1} for  $0\leq A<\frac{1}{2}$  we obtain
\begin{eqnarray*} 
&  &
(\tau +1) \left((x+1)^2-A^2\right)  F \left(-\frac{1}{2},-\frac{1}{2};1;\frac{(1-\tau )^2-A^2}{(x+1)^2-A^2}\right)\\
&  &
- \left(-A^2-\tau +1\right)  F \left( \frac{1}{2}, \frac{1}{2};2;\frac{(1-\tau )^2-A^2}{(\tau +1)^2-A^2}\right)\\
& = &
(\tau +1) \left((\tau +1)^2-A^2\right)   \left(\frac{ \Gamma (-2k)}{[\Gamma (\frac{1}{2}-k)]^2}-z \left(k+\frac{1}{2}\right)^2  \frac{ \Gamma (-1-2k)}{[\Gamma (\frac{1}{2}-k)]^2}+O(z^2)\right)\\
&  &
- \left( 1-A^2-\tau \right)   \left( \frac{ 4}{ \pi} +(-z)  \Big\{ 
\frac{1}{  {\pi } }  
\left[ 2\psi ( 1) - 2\psi (\frac{3}{2} ) - \ln ( z) \right]   
+O(1) ( z) \ln ( z)\Big\} \right)\\
& = &
(\tau +1) \left((\tau +1)^2-A^2\right)   \left(\frac{ 4}{\pi}-\left( \frac{4x}{\left(1+\tau \right)^2-A^2}\right)\left( -\frac{1}{2}\right)^2  \frac{ 4}{ \pi}+O(z^2)\right)\\
&  &
- \left( 1-A^2-\tau \right)   \left( \frac{4}{ \pi}  -\left( \frac{4x}{\left(1+x\right)^2-A^2}\right) \left\{ 
\frac{1}{  {\pi } }  
\left[ 2\psi ( 1) - 2\psi (\frac{3}{2} ) - \ln ( z) \right]   
+O(1) ( z) \ln ( z)\right\} \right)\\
&= &
- \tau \frac{4}{\pi } \left(H^2 r^2-\ln \left(4-4 H^2 r^2\right)+\ln (\tau )+1\right)+ O\left(\tau ^2\right)\\
& = &
 - \tau  \frac{4}{\pi } \left(  \ln (\tau )+ O(1)\right)+O\left(\tau ^2\right) \quad 
\mbox{\rm as }\quad t \rightarrow  \infty \quad 
\mbox{\rm and }\quad \tau =e^{-Ht} \rightarrow 0.
\end{eqnarray*}
\medskip

\noindent
{\bf The proof of (i) Theorem~\ref{T0.3} for the case of $m=-iH/2$, $\ell=-2 $, $k=-1$.  }
We have
\begin{eqnarray*} 
\Psi_1 (0,t) 
& = &
   2^2 e^{-Ht}  
  \int_{0}^{1 } \left( \frac{\partial }{\partial r} r\Phi _0 ( r) \right) \Bigg\{ -\frac{H }{8 \left(\left(e^{-H t}+1\right)^2-H^2 r^2\right)^{3/2}}\\
&  &
\times  \Bigg[  \left(e^{-H t}+1\right) \left(\left(e^{-H t}+1\right)^2-H^2 r^2\right) F \left(-\frac{1}{2},-\frac{1}{2};1;\frac{\left(-1+e^{-H t}\right)^2-H^2 r^2}{\left(1+e^{-H t}\right)^2-H^2 r^2}\right)\\
&  &
+  \left(H^2 r^2+e^{-2 H t}-1\right)  F \left(\frac{1}{2},\frac{1}{2};2;\frac{\left(-1+e^{-H t}\right)^2-H^2 r^2}{\left(1+e^{-H t}\right)^2-H^2 r^2}\right)  \Bigg]\Bigg\} dr  \\ 
& = &
   2^2 e^{-Ht}  
  \int_{0}^{1 } \left( \frac{\partial }{\partial r} r\Phi _0 ( r) \right) \Bigg\{ -\frac{H }{8 \left(\left(e^{-H t}+1\right)^2-H^2 r^2\right)^{3/2}}\\
&  &
\times  \Bigg[ -\frac{4 e^{-H t} \left(H^2 r^2-\log \left(4-4 H^2 r^2\right)-Ht+1\right)}{\pi }+e^{-2H t}O\left(1\right) \Bigg]\Bigg\} dr  \,. 
\end{eqnarray*}
Finally,
\begin{eqnarray*} 
\Psi_1 (0,t) 
& = & 
 e^{-2Ht}  H 
  \int_{0}^{1 } \left( \frac{\partial }{\partial r} r\Phi _0 ( r) \right)   \frac{1}{2 \left(\left(e^{-H t}+1\right)^2-H^2 r^2\right)^{3/2}} \\
&  &
\times  \Bigg[  \frac{4}{\pi }  \left(H^2 r^2-\log \left(4-4 H^2 r^2\right)-Ht+1\right)+e^{- H t}O\left(1\right) \Bigg]  dr \,. 
\end{eqnarray*}
Next we choose $\Phi _0 $ such that
\begin{eqnarray*} 
&   & 
  \int_{0}^{1 } \left( \frac{\partial }{\partial r} r\Phi _0 ( r) \right)   \frac{1}{  \left(1-H^2 r^2\right)^{3/2}} 
\Bigg[  \frac{4}{\pi }  \left(H^2 r^2-\log \left(4-4 H^2 r^2\right)-Ht+1\right) \Bigg]  dr  \not= 0\,.
\end{eqnarray*}
The last equation shows that the value  $\Psi_1 (0,t)$ for large time depends on the values of initial function inside of the characteristic conoid. This completes the proof in this case. 

\section {{\bf Necessity of $m=0,\pm iH$  for the Huygens  principle.  Case of $ m = i\frac{H}{2} $, $\ell=0$}} 
\label{S7}

For   $m= i\frac{H}{2}$ we have $ M_+=0$ and 
\begin{eqnarray*}
&  &
\left(   \frac{\partial}{\partial t} - \frac{H}{2}  -im\right)  
 K_1( r,t;M_+)   \\
&  = &
H  e^{-  H t} \left(\left(1+e^{-H t}\right)^2-H^2 r^2\right)^{-\frac{5}{2}} \\
&  &
\times \Bigg\{\left(1+e^{-H t} \right) \left(\left(1+e^{-H t} \right)^2-H^2 r^2\right)  
F \left(\frac{ 1}{2},\frac{ 1}{2};1;\frac{\left(1-e^{-H t}\right)^2-H^2 r^2}{\left(1+e^{-H t}\right)^2-H^2 r^2}\right)\\
&  &
+\left(1-e^{-2 H t}-H^2 r^2 \right)  
F \left(\frac{3}{2},\frac{3}{2};2;\frac{\left(1-e^{-H t}\right)^2-H^2 r^2}{\left(1+e^{-H t}\right)^2-H^2 r^2}\right)\Bigg\}\,.
\end{eqnarray*}
We are going to study the asymptotics of the function  
\[
\left(1+x \right) \left(\left(1+x \right)^2-H^2 r^2\right)  
F \left(\frac{ 1}{2},\frac{ 1}{2};1;\frac{\left(1-x\right)^2-A^2}{\left(1+x\right)^2-A^2}\right) 
+\left(1-x^{2  }-A^2 \right)  
F \left(\frac{3}{2},\frac{3}{2};2;\frac{\left(1-x\right)^2-A^2}{\left(1+x\right)^2-A^2}\right),
\]
where $\tau =e^{-H t} $ and $A= Hr $ as $ \tau  \to 0$.
By (\ref{A2}) with $a=b=1/2$ and $m=0$ we obtain
\begin{eqnarray*}
F \left(\frac{ 1}{2},\frac{ 1}{2};1;1-z\right)  
& = &
\frac{1}{\pi}   
\left[ 2\psi (n+1) - 2\psi \left(\frac{ 1}{2}\right)  - \ln ( z) \right]\\
&  &
+\frac{1}{\pi}\sum _{n=1}^\infty \frac{(a)_n(b)_n}{(n!)^2}
\left[ 2\psi (n+1) - \psi (a+n)- \psi (b+n) - \ln ( z) \right]   z ^{n} \\
& = &
\frac{1}{\pi}   
\left[ 2\psi (n+1) - 2\psi \left(\frac{ 1}{2}\right)  - \ln ( z) \right] +z\ln (z)O(1)\quad \mbox{\rm as } \,\, z \searrow 0\,.
\end{eqnarray*}
For the second hypergeometric    function,  
 since $c-a-b=-1$, we apply (\ref{mBEA})  with $m=1$:
\begin{eqnarray*} 
F \left(\frac{3}{2},\frac{3}{2};2;1- z \right)  
&= & 
\frac{ 4z^{-1}}{ \pi }    
-\frac{1}{\pi} \sum_{n=0}^{\infty} \frac{[(\frac{3}{2})_{n}]^2}{(n+1)_{n} n !}\left[\bar{h}_{n}-\ln ( z)\right]  z ^{n} \\ 
&= & 
\frac{ 4}{ \pi }z^{-1}    
-O(1) \ln (z) \quad \mbox{\rm as } \,\, z \searrow 0\,,
\end{eqnarray*} 
where  
$\bar{h}_{n}=\psi(1+n)+\psi(2+n )-\psi(a+n)-\psi(b+n)$. 
Hence
\begin{eqnarray*}
&  &
\left(1+\tau  \right) \left(\left(1+\tau  \right)^2-A^2\right)  
F \left(\frac{ 1}{2},\frac{ 1}{2};1;\frac{\left(1-\tau \right)^2-A^2}{\left(1+\tau \right)^2-A^2}\right)\\
&  &
+\left(1-\tau ^{2  }-A^2 \right)  
F \left(\frac{3}{2},\frac{3}{2};2;\frac{\left(1-\tau \right)^2-A^2}{\left(1+\tau \right)^2-A^2}\right)\\
& = &
\left(1+\tau  \right) \left(\left(1+\tau  \right)^2-A^2\right)  
  \left(\frac{1}{\pi}   
\left[ 2\psi (n+1) - 2\psi (\frac{ 1}{2})  - \ln ( z) \right] +z\ln (z)O(1)\right)\\
&  &
+\left(1-\tau ^{2  }-A^2 \right)  
 \left(\frac{ 4z^{-1}}{ \pi }    
-O(1)\ln (z) \right)\\
& = &
\left(1-A^2 \right)  
 \left(\frac{ 4}{ \pi }\left(\frac{4e^{-H t} }{1-A^2}\right)^{-1}    
-O(1)\ln \left(\frac{4e^{-H t} }{1-A^2}\right) \right)\\
& = &
\left(1-A^2 \right)  
 \left(  \frac{ 1}{ \pi }  \left( 1-A^2 \right)   e^{H t}  
-O(1)t \right)\\
& = &
\left(1-A^2 \right)^2  
    \frac{ 1}{ \pi }     e^{H t}  
-O(1)t
\end{eqnarray*}
and, consequently, 
\begin{eqnarray*} 
\left(   \frac{\partial}{\partial t} - \frac{H}{2}  -im\right)  
 K_1( r,t;M_+)   
& = &\frac{H^3}{\pi  \sqrt{1-H^2 r^2}}+e^{-2H t}O\left(1\right)\,.
\end{eqnarray*}
\medskip

\noindent
{\bf The proof of necessity part of Theorem~\ref{T0.3} for the case of $m=iH/2$.  }
For the solution for large time we obtain
\begin{eqnarray*}
\Psi_1 (0,t) 
& = &
  2 e^{-Ht}  
  \int_{0}^{1 } \left( \frac{\partial }{\partial r} r\Phi _0 ( r) \right) \left(   \frac{\partial}{\partial t}  \right)  
 K_1( r,t;0)  \,  dr  \\ 
& = &
  2 e^{-Ht}  
  \int_{0}^{1 } \left( \frac{\partial }{\partial r} r\Phi _0 ( r) \right) \left(  \frac{H }{\pi  \sqrt{1-H^2 r^2}}+e^{-2H t}O\left(1\right) \right) \,  
   dr \,.
\end{eqnarray*}
Next we choose $ \Phi _0$ such that
\begin{eqnarray*}   
  \int_{0}^{1 } \left( \frac{\partial }{\partial r} r\Phi _0 ( r) \right)    \frac{1}{  \sqrt{1-H^2 r^2}}      dr \not= 0\,.
\end{eqnarray*}
The last equation shows that the value  $\Psi_1 (0,t)$ for large time depends on the values of initial function inside of the characteristic conoid. This completes the proof in this case.

\section{Necessity of $m=\pm iH $ for the incomplete Huygens  principle}

In order to prove necessity of $m= iH $ for the incomplete Huygens  principle with respect to the first 2-spinor  initial data $\Phi _0, \Phi _1 $ for the Dirac equation (\ref{DE}) we set $m \not=0$, $m\not= iH$ and chose initial data  
(\ref{3.27}). Then we can repeat all arguments used for the cases of all possible values of mass except ones used in Sections~\ref{S6}, \ref{S7}. This completes the proof of necessity part in   Theorem~\ref{T0.3} for $m=iH$.   

In order to prove  necessity of  $m= -iH $  for the incomplete Huygens  principle with respect to the second 2-spinor  initial data $\Phi _2, \Phi _3 $ for the Dirac equation (\ref{DE}) we set $m \not=0$, $m\not= -iH$ and 
 choose initial data  (\ref{3.21}) 
as in subsection~\ref{S4}.
We set $F=0$ and $M_-=\frac{H}{2}-im \not= - \frac{H}{2}$, then the kernel function is  
\begin{eqnarray*} 
K_1(r,t;M_-)
& =  &
  2^{-1+\frac{2 i m}{H}} e^{\frac{1}{2} t (H-2 i m)} \left(\left(e^{-H t}+1\right)^2-H^2 r^2\right)^{-\frac{i m}{H}} \\
&  &
\times  F \left(\frac{i m}{H},\frac{i m}{H};1;\frac{\left(-1+e^{-H t}\right)^2-H^2 r^2}{\left(1+e^{-H t}\right)^2-H^2 r^2}\right)\,, \nonumber
\end{eqnarray*} 
while the 
operator $ {\cal K}_1(x,t,D_x;M_-)$ is defined by (\ref{K1OPER}).  The solution to the Cauchy problem for the Dirac equation is the function
\begin{eqnarray*} 
\Psi (x,t) 
& = &
  e^{-Ht}\left(   \partial_0 {\mathbb I}_4+  e^{-Ht} \gamma ^k \gamma ^0\partial_k- \frac{H}{2}{\mathbb I}_4 -im\gamma ^0\right)   \left (
   \begin{array}{cccc}
0  \\
0 \\
 {\cal K}_1(x,t,D_x;M_-)[\Phi _2(x )]  \\
0\\
   \end{array}
   \right ) \,,
\end{eqnarray*}
with the components (see (\ref{5.29}))
\begin{eqnarray*} 
\Psi_0 (x,t) 
& = &
   - e^{-2Ht}  \partial_3    
 {\cal K}_1(x,t,D_x;M_-)[\Phi _2(x ) ] \,, \\
 \Psi_1 (x,t) 
& = &
  e^{-2Ht}\left(  -   \partial_1-i  \partial_2   \right){\cal K}_1(x,t,D_x;M_-)[\Phi _2(x ) ]
\,,\\
 \Psi_2 (x,t) 
& = &
  e^{-Ht}\left( \partial_0-   \frac{H}{2}  +im \right)  {\cal K}_1(x,t,D_x;M_-)[\Phi _2(x ) ]\,,\\
 \Psi_3 (x,t) 
& = &
 0\,.
\end{eqnarray*} 
We calculate
\begin{eqnarray*} 
\Psi_2 (x,t) 
& = &
  e^{-Ht}\left(  \frac{\partial}{\partial t} - \frac{H}{2}  +im\right)   
 \, 2\int_{0}^{\phi (t) }  v_{\Phi _2 } (x,  s)
  K_1( s,t;M_-)   ds  \,.
\end{eqnarray*}
It  can be rewritten    in the terms of the function $V_{\Phi _2} $ defined in accordance to (\ref{14}), that is, 
\begin{eqnarray*} 
\Psi_2 (x,t) 
& = &
  e^{-Ht}\left(   \frac{\partial}{\partial t} - \frac{H}{2}  +im\right)   
 \, 2\int_{0}^{\phi (t) } \left( \frac{\partial }{\partial s}V_{\Phi _2}(x,s)  \right)
  K_1( s,t;M_-)   ds \,.
\end{eqnarray*}
It follows,
\begin{eqnarray*} 
\Psi_2 (x,t) 
& = &
  e^{-Ht}\left(   \frac{\partial}{\partial t} - \frac{H}{2}  +im\right)   
 \, 2 \Big(  V_{\Phi _0}(x, \phi (t))  K_1(\phi (t),t;M_-) -  V_{\Phi _2}(x, 0)  K_1(0,t;M_-)
 \Big)  \\
&   &
 - e^{-Ht}\left(   \frac{\partial}{\partial t} - \frac{H}{2}  +im\right)   
 \, 2\int_{0}^{\phi (t) } V_{\Phi _0}(x,s)  
 \frac{\partial }{\partial s} K_1( s,t;M_-)   ds \\
& = &
 2   e^{-Ht}\left(   \frac{\partial}{\partial t} - \frac{H}{2}  +im\right)   
    V_{\Phi _2}(x, \phi (t))  K_1(\phi (t),t;M_-)   \\
&   &
 - 2 e^{-2Ht}   
   V_{\Phi _2}(x,\phi (t))   
\left(  \frac{\partial  }{\partial s  } K_1( s,t;M_-) \right)_{s=\phi (t)}  \\
&   &
 - 2 e^{-Ht}  
  \int_{0}^{\phi (t) } V_{\Phi _2}(x,s) \left(   \frac{\partial}{\partial t} - \frac{H}{2} +im\right)  
 \frac{\partial }{\partial s} K_1( s,t;M_-)   ds \,.
\end{eqnarray*} 
In particular, since $x \in {\mathbb R}^3$, by the Kirchhoff's formula we have 
\begin{eqnarray*}
V_{\Phi _2}(0, \phi (t))=\phi (t) \Phi _2 (\phi (t)) =\frac{ 1-e^{-Ht} }{H} \Phi _2 \left(\frac{ 1-e^{-Ht} }{H} \right)=0 
\end{eqnarray*} 
for sufficiently large $t$, that is, if $  1-e^{-Ht}  > H  \varepsilon   $. Consequently, for large $t$ we have 
\begin{eqnarray*} 
\Psi_2 (0,t) 
& = &
 - 2 e^{-Ht}  
  \int_{0}^{\phi (t) } s\Phi _0 ( s) \left(   \frac{\partial}{\partial t} - \frac{H}{2}  +im\right)  
 \frac{\partial }{\partial s} K_1( s,t;M_-)   ds  \nonumber \\ 
& = &
 - 2 e^{-Ht}  
  \int_{0}^{1 } s\Phi _0 ( s) \left(   \frac{\partial}{\partial t} - \frac{H}{2} +im\right)  
 \frac{\partial }{\partial s} K_1( s,t;M_-)   ds  \nonumber \\ 
& = &
   2 e^{-Ht}  
  \int_{0}^{1 } \left( \frac{\partial }{\partial s} s\Phi _0 ( s) \right) \left(   \frac{\partial}{\partial t} - \frac{H}{2} +im\right)  
 K_1( s,t;M_-)   ds   \,.
\end{eqnarray*} 
Now we focus on the tail of the solution, that is on the term  generated by the  integral. To discuss the last term we apply Lemma~\ref{L4.1} when $m$ is replaced with $-m$. Further,  
according to Proposition~~\ref{P4.2} if  $m \in {\mathbb C}$ and 
\[
 m \not= i\frac{H}{2}   +  i\frac{H}{2}\ell, \quad  \ell=0,\pm 1, \pm2,\ldots \,,
\] 
then
\begin{eqnarray*} 
&  &
2\frac{ i m}{H}  \left( 1 -e^{- 2H t} - H^2 r^2  \right) 
 F \left(1+\frac{i m}{H} ,1+\frac{i m}{H} ;2;\frac{\left( 1-e^{-H t}\right)^2-H^2 r^2}{\left(1+e^{-H t}\right)^2-H^2 r^2}\right) \\
&  &
+   \left(1+e^{-H t} \right) \left(  (1 + e^{-H t})^2-H^2 r^2\right)
 F \left(\frac{i m}{H},\frac{i m}{H};1;\frac{\left( 1-e^{-H t}\right)^2-H^2 r^2}{\left(1+e^{-H t}\right)^2-H^2 r^2}\right) \\
&  = & 
-2 \frac{im }{H   }   4 ^{-2\frac{i m}{H} }   e^{ 2 i m   t}   
\frac{  \Gamma( 2\frac{i m}{H} )}{[\Gamma(1+\frac{i m}{H})]^2 }\left(  1-  H^2 r^2\right)^{1+2\frac{i m}{H}}  
   +  R (m,H,r;t)   \,,
\end{eqnarray*} 
where  with  large  $T$  the remainder  $R (m,H,r;t) $    can be estimated by (\ref{4.25}). Further,
\begin{eqnarray*} 
\Psi_2 (0,t) 
& = &
   2 e^{-Ht}  
  \int_{0}^{1 } \left( \frac{\partial }{\partial r} r\Phi _0 ( r) \right) \Bigg[   
 2^{\frac{2 i m}{H}}i m   e^{\frac{1}{2} t (-H-2 i m)} \left(\left(1+e^{-H t} \right)^2-H^2 r^2\right)^{-\frac{i m}{H}-2}  \\
&  &
\times \Bigg\{  2\frac{ i m}{H}  \left( 1 -e^{- 2H t} - H^2 r^2  \right) 
 F \left(1+\frac{i m}{H} ,1+\frac{i m}{H} ;2;\frac{\left( 1-e^{-H t}\right)^2-H^2 r^2}{\left(1+e^{-H t}\right)^2-H^2 r^2}\right) \\
&  &
+   \left(1+e^{-H t} \right) \left(  (1 + e^{-H t})^2-H^2 r^2\right)
 F \left(\frac{i m}{H},\frac{i m}{H};1;\frac{\left( 1-e^{-H t}\right)^2-H^2 r^2}{\left(1+e^{-H t}\right)^2-H^2 r^2}\right) \Bigg\}\Bigg]     dr \\ 
& = &
   2 e^{-Ht}  2^{\frac{2 i m}{H}}i m   e^{\frac{1}{2} t (-H-2 i m)}
  \int_{0}^{1 } \left( \frac{\partial }{\partial r} r\Phi _0 ( r) \right)   
 \left(\left(1+e^{-H t} \right)^2-H^2 r^2\right)^{-\frac{i m}{H}-2}  \\
&  &
\times \Bigg\{ -2 \frac{im }{H   }   4 ^{-2\frac{i m}{H} }   e^{ 2 i m   t}   
\frac{  \Gamma( 2\frac{i m}{H} )}{[\Gamma(1+\frac{i m}{H})]^2 }\left(  1-  H^2 r^2\right)^{1+2\frac{i m}{H}}  
   +  R (m,H,r;t) \Bigg\}     dr  \,.
\end{eqnarray*} 
Since $ -1+ \frac{i m}{H}\not= 0$ we  can choose the radial function $ \Phi _0 \in C^\infty (B_\varepsilon (0)) $ such that
\begin{eqnarray*}   
&  &
  \int_{0}^{1 } \left( \frac{\partial }{\partial r} r\Phi _0 ( r) \right) 
\left(  1-  H^2 r^2\right)^{-1+ \frac{i m}{H}}  
    dr\not= 0\,.
\end{eqnarray*}
The last equation shows that the value  $\Psi_2 (0,t)$ for large time depends on the values of initial function inside of the characteristic conoid. This completes the proof in this case. \hfill $\square$  
\medskip 

\section{Conclusions}

The purpose of this paper was to examine the  Huygens  principle
for generalized Dirac operator in the de~Sitter spacetime. The generalized Dirac operator was firstly introduced in \cite{ArX2020} and besides the  mathematical importance for the theory of partial differential equations,  it, in particular, includes the equation for the motion of the charged spin-$\frac{1}{2}$ particle in a 
constant homogeneous magnetic field. The last problem  has been studied in physical literature (see \cite{ArX2020} and the references therein).   
In the present paper, we  introduced a novel definition of the so-called incomplete Huygens  principle for this operator and comprehensively  examined it. The incomplete Huygens  principle was introduced in \cite{JMP2013} for the scalar fields satisfying  the Klein-Gordon equation in the de~Sitter spacetime. In  \cite{JMP2013} it was   shown that an existence of two scalar fields obeying the incomplete Huygens  principle  in the de~Sitter spacetime implies that  the spacetime is four-dimensional. The  usual splitting of four-spinors into 2-spinors  allowed us to verify  the  Huygens  principle for each 2-spinor separately. Mathematical analysis of the explicit formulas for the solutions, which is summarized in the main result of the present paper,  Theorem~\ref{T0.3},   led  to three exceptional values for the mass of the field: $m=0$, $m=-iHh/c^2$, and $m=iHh/c^2$, where $H $ is the Hubble constant,  $c$ is the speed of light, and  $h$ is Planck's constant. It is remarkable that the duration of time  when the factor $\exp \left(   - i \,m c^2 t/h\right)  $ of the kernel 
of the integral representing   the solution  of the generalized Dirac equation  with the mass $m=\pm i Hh/c^2$ stays around unity,   is limited by  $ \approx  10^{18}$ sec, which coincides with the age of the universe. A short paragraph written in the paper about neutrinos and imaginary mass touched on the possible physical interpretation in the framework of the Standard Model of particles physics. A more comprehensive  discussion  of the physical interpretation of the imaginary mass appearing as a result of mathematical analysis of the   generalized Dirac equation in the de~Sitter spacetime including, for instance,  tachyons, was out of the scope of the present paper.

\begin{appendix}
\newtheorem{lemmaA}[theorem]{Lemma A}

\renewcommand{\theequation}{\thesection.\arabic{equation}}
\setcounter{equation}{0}

\section{Appendix}

\subsection{Some properties of hypergeometric function}

There is a 
formula (See 15.3.6 of Ch.15\cite{A-S}  and \cite[Sec.2.3.1]{B-E}.) 
that ties together points $z=0$ and $z=1$:
\begin{eqnarray} 
\label{15.3.6A}
 F  \left( a,b;c;z  \right) 
& = &
\frac{\Gamma (c)\Gamma (c-a-b)}{\Gamma (c-a)\Gamma (c-b)}F  \left( a,b;a+b-c+1;1-z  \right) \\
 &  &
 + (1-z)^{c-a-b}\frac{\Gamma (c)\Gamma (a+b-c)}{\Gamma (a)\Gamma (b)}F  \left( c-a,c-b;c-a-b+1;1-z  \right) \,, \nonumber  
\end{eqnarray}
where $|\arg (1-z)| <\pi$, $\quad |1-z|<1 $,  
and  $c-a-b \not= \pm 1,\pm 2,\ldots\,$. 

\subsection{Some properties of hypergeometric function. Case of zeros}

We use (12) of \cite[Sec. 2.10]{B-E}:
\begin{eqnarray}
\label{A2}
F(a, b ; a+b+m ; z) \frac{1}{\Gamma(a+b+m)}
& = &
\frac{\Gamma(m)}{\Gamma(a+m) \Gamma(b+m)} \sum_{n=0}^{m-1} \frac{(a)_{n}(b)_{n}}{(1-m)_{n} n !}(1-z)^{n} \\
&  &
+\frac{(1-z)^{m}(-1)^{m}}{\Gamma(a) \Gamma(b)} \sum_{n=0}^{\infty} \frac{(a+m)_{n}(b+m)_{n}}{(n+m) ! n !}\left[h_{n}^{\prime \prime}-\ln (1-z)\right](1-z)^{n} \,,  \nonumber  
\end{eqnarray}
where  $
-\pi<\arg (1-z)<x$, \quad $a, b, \neq 0,-1,2, \ldots $,    
\[
h_{n}^{\prime \prime}=\psi(n+1)+\psi(n+m+1)-\psi(a+n+m)-\psi(b+n+m), \nonumber
\]
and $\sum_{n}^{m-1}$ is to be interpreted as zero when $m=0$.    
The function $\psi(z)$ is the logarithmic derivative of the gamma function:
$
 \psi(z)=\frac{d \ln \Gamma(z)}{d z}=\frac{\Gamma^{\prime}(z)}{\Gamma(z)}$.

\subsection{Some properties of hypergeometric function. Case of poles}
\label{ssA3}

If $k=1,2,3,\ldots$ since $c-a-b=-2k$  then we apply \cite[(14) Sec. 2.10]{B-E}:
\begin{eqnarray}
\label{mBEA}
F(a, b, a+b-m ; z) \frac{1}{\Gamma(a+b-m)} 
&= & 
\frac{\Gamma(m)(1-z)^{-m}}{\Gamma(a) \Gamma(b)} \sum_{n=0}^{m-1} \frac{(a-m)_{n}(b-m)_{n}}{(1-m)_{n} n !}(1-z)^{n}    \\
&  &
+\frac{(-1)^ m}{\Gamma(a-m) \Gamma(b-m)} \sum_{n=0}^{\infty} \frac{(a)_{n}(b)_{n}}{(n+m)_{n} n !}\left[\bar{h}_{n}-\ln (1-z)\right](1-z)^{n} \,, \nonumber
\end{eqnarray} 
where  
$-\pi<\arg (1-z)<\pi, \quad a, b, \neq 0,-1,-2, \dots
$
\[
\bar{h}_{n}=\psi(1+n)+\psi(1+n+m)-\psi(a+n)-\psi(b+n)\,,
\]
and  $\sum_{n=0}^{m-1} $ is set zero if  $m=0$.

\end{appendix}

\end{document}